\documentclass[draft]{agujournal2019}
\usepackage{url}
\usepackage{amsmath}
\usepackage{lineno}
\usepackage[inline]{trackchanges}
\usepackage{soul}
\usepackage{placeins}
\usepackage{color}
\usepackage{subcaption}
\providecommand{\citep}{\cite}
\providecommand{\citet}{\citeA}
\draftfalse

\journalname{Journal of Earth Sciences}

\begin{document}
\justifying

\title{The 2016  $M_w$  7.0 Kumamoto Earthquake Sequence, Japan revisited: Insights from Spatio-Temporal Analysis of Seismicity Parameters}

\authors{Hamzeh Mohammadigheymasi\affil{1,2}, Muhammed Hossein Mousavi\affil{3},  Marjan Tourani\affil{4}, and Nasrin Tavakolizadeh\affil{2}}

\affiliation{1}{Atmosphere and Ocean Research Institute (AORI), the University of Tokyo, Kashiwa, Japan}
\affiliation{2}{Department of Computer Sciences, University of Beira Interior, Covilha, Portugal}
\affiliation{3}{Department of Physics Education, Farhangian University, P.O. Box 14665-889, Tehran, Iran}
\affiliation{4}{Department of Geological Engineering, Tectonics Research Group, Ankara University, Ankara, Türkiye}

\correspondingauthor{Hamzeh Mohammadigheymasi}{hamzeh@ubi.pt}

\begin{keypoints}
	\item Joint mapping of the $b$-value, fractal dimension ($D_c$), and $Z$-value reveals an acute preparatory phase of shear-stress concentration and localized fault locking prior to the 2016 Kumamoto mainshock.
	\item Coseismic rupture rapidly reversed these precursory signals, driving a surge in $b$-value and $D_c$ that reflects a significant stress drop and volumetric aftershock activation.
	\item Post-seismic depth cross-sections identify discrete, unrelaxed residual asperities at the southwestern Hinagu termination and near Mount Aso, consistent with localized Coulomb stress loading.
\end{keypoints}

\begin{abstract}
	Understanding how crustal fault systems accumulate tectonic strain, nucleate catastrophic ruptures, and redistribute post-seismic stress is fundamental to earthquake physics and seismic hazard assessment. The 16 April 2016 $M_w$ 7.0 Kumamoto earthquake sequence, which ruptured the multi-segment Futagawa--Hinagu fault zone within the Beppu--Shimabara graben in central Kyushu, Japan, offers a premier dataset for tracking these processes across a complete earthquake cycle. Integrating the dense Japan Meteorological Agency catalog spanning 2014--2018 with dynamic completeness estimation ($M_c \approx 2.15$--$2.25$), we jointly examine the spatio-temporal evolution of three complementary seismicity parameters: the Gutenberg--Richter $b$-value, the three-dimensional hypocentral fractal dimension ($D_c$), and the standardized seismicity-rate anomaly ($Z$-value). During the one- to two-year preparatory phase, the regional $b$-value declined systematically from a background baseline of $b \approx 1.35 \pm 0.10$ to an anticipatory precursory minimum of $b \approx 0.59$, while $D_c$ contracted from $0.85 \pm 0.15$ to $0.70$--$0.80$, jointly recording shear-stress concentration and progressive microfracture coalescence onto a narrow, fault-parallel nucleation zone. A spatially coherent negative $Z$-value anomaly ($Z \approx -1.6$ to $-2.0$) developed concurrently along the Beppu--Shimabara graben and strengthened with increasing integration time, approaching the adopted significance threshold only in the longest (five-month) test window; taken together with the $b$- and $D_c$-value trends, this anomaly is consistent with, though not independent proof of, progressive fault locking. Pseudo-tomographic three-dimensional depth-sliced volumes and consecutive four-month spatio-temporal stacks show that this low-$b$ locked core ($b \le 0.65$) was vertically stratified at $10.0$--$12.5$~km depth and developed acutely within the final four months (February--April 2016) preceding failure, rather than as a static, long-lived feature. Coseismic rupture reversed this state within weeks: the $b$-value surged to $1.25$--$1.35$ and $D_c$ expanded to $2.04$--$2.11$, consistent with the coseismic stress drop ($\Delta\tau \sim 1$--$10$~MPa) and volumetric aftershock activation across conjugate fault branches, followed by gradual recovery toward background levels over the subsequent 2.5 years. Fault-parallel and transverse depth cross-sections ($A_1$--$B_1$, $A_2$--$B_2$, $A_3$--$B_3$) reveal sharp planar aftershock localization ($D_c \approx 1.20$--$1.35$) alongside two discrete, unrelaxed residual asperities ($b \approx 0.80$--$0.95$) at the southwestern Hinagu termination and adjacent to the Mount Aso volcanic conduit, consistent with localized positive Coulomb stress loading ($\Delta\mathrm{CFS} > 0$). This joint multi-parameter framework resolves the complete lifecycle of asperity locking, dynamic release, and heterogeneous post-seismic healing more diagnostically than any single metric alone; however, because the reported anomalies were identified retrospectively with foreknowledge of the mainshock, they should be read as evidence of coherent crustal mechanical behavior rather than as a validated operational forecasting tool.
\end{abstract}

\section{Introduction}\label{sec:intro}
Earthquakes are among the most devastating natural phenomena, causing widespread infrastructural collapse, disrupting essential utility networks, and generating profound socioeconomic crises in seismically active regions worldwide \citep{coburn2002earthquake, cui2011}. Over the past century, a series of massive megathrust ruptures, including the 1960 $M_w$ 9.5 Valdivia (Chile), 1964 $M_w$ 9.2 Prince William Sound (Alaska), 2004 $M_w$ 9.1--9.3 Sumatra, and 2011 $M_w$ 9.0 T\=ohoku-Oki (Japan) earthquakes, have exposed the vulnerability of modern civilization to sudden subduction-zone fault slip \citep{cifuentes1989chile, plafker1965tectonic, kazama2012, nanjo2012}. Shallow crustal strike-slip earthquakes pose a comparable hazard despite their smaller magnitudes, owing to their proximity to populated areas and the intensity of near-source ground shaking; the 2023 $M_w$ 7.8 Kahramanmara\c{s} (T\"urkiye) doublet and the 2016 $M_w$ 7.0 Kumamoto (Japan) earthquake are recent examples of this class of destructive events.

To advance earthquake hazard mitigation and monitor the mechanical state of such active fault systems, statistical seismology has emerged as a critical framework that transforms high-resolution earthquake catalogs \citep{mohammadigheymasi2023ipiml, carvalho2025application, khurshid2025comprehensive, mohammadigheymasi2023application} into active, in situ probes of crustal behavior. Because direct geomechanical measurements at seismogenic depths of 10 to 20~km remain extremely difficult to obtain, researchers increasingly rely on indirect differential stress evaluation methods to estimate physical conditions within active fault zones. Wiemer and Wyss established the operational framework for extracting such proxies from routine catalogs, first through systematic mapping of spatial $b$-value variability \citep{wiemer2000} and subsequently through a dedicated review of quiescence and precursory rate change detection \citep{wiemer2002}; together, these studies define the two principal observational axes, namely spatial heterogeneity and temporal rate change, that the present work builds upon. In practice, spatial variations in seismicity parameters resolve highly stressed locked asperities from stress-released zones. This physical link was first established experimentally by \citet{scholz1968}, later confirmed through systematic catalog analysis by \citet{wiemer1997}, and subsequently formalized into a framework linking $b$-value heterogeneity to fault segmentation by \citet{schorlemmer2005}. Spatial clustering additionally quantifies the structural complexity of the active fracture network via the fractal dimension \citep{hirata1989}, while tracking these parameters through time distinguishes mechanical stress-driven processes from fluid-driven anomalies such as pore-pressure diffusion within hydrothermal systems \citep{miller2004, fischer2014, farrell2009, hill2003, pacchiani2010}. Developing robust differential stress estimation methods that integrate these diverse indicators therefore remains a central challenge in linking statistical seismicity modeling to physical models of crustal loading and seismic hazard evaluation.

An exceptional target for applying such an integrated differential stress evaluation framework is the 2016 Kumamoto earthquake sequence in central Kyushu, southwestern Japan. The sequence was characterized by a major strike-slip event that ruptured the Futagawa--Hinagu fault zone, part of the broader right-lateral shear system associated with the Median Tectonic Line and the Beppu--Shimabara graben \citep{Kato2016}. The mainshock nucleated on April 15, 2016, at 16:25 GMT (April 16 at 01:25 local time), with an epicenter near Mashiki, Kumamoto Prefecture ($32.754^\circ\text{N}$, $130.763^\circ\text{E}$), at a shallow crustal depth of 12.5~km \citep{Kato2016}. It was preceded by a sequence of strong foreshocks on the same fault system, including $M_j$~6.5 (April 14, ${\sim}21{:}26$ JST) and $M_j$~6.4 (April 15, ${\sim}00{:}03$ JST) events---$M_j$ denoting the Japan Meteorological Agency (JMA) magnitude scale routinely used for these regional events---and was followed by numerous aftershocks, several exceeding $M_j$~5 \citep{Asano_2016, Kubo_2016, Yoshida_2016, Kato_2016, Wang_2019}. The sequence produced intense ground shaking and severe damage across Kumamoto Prefecture and adjacent regions, with official tallies reporting over 50 direct fatalities (rising to more than 270 when disaster-related deaths are included) and over 1,000 injuries \citep{Wang_2019, Scott_2019}.

As with the 2016 Kumamoto sequence, most major earthquakes are accompanied by foreshock and aftershock activity that records the evolving stress field, fault interaction, and crustal deformation associated with the rupture process \cite<e.g., >{bayrak2013, Thapa_2018, Wang_2019, mousavi2025b,mousavi2026cross}. Statistical analysis of such sequences has therefore become a central tool in seismological research. Among the most widely applied approaches are the Gutenberg--Richter frequency--magnitude relation ($b$-value), the fractal dimension ($D_c$-value), and the seismic quiescence parameter ($Z$-value), which together characterize the magnitude distribution, spatial clustering, and temporal rate changes of seismicity, offering insight into stress evolution and fault behavior during earthquake sequences \cite<e.g., >{Ogata_1999, Telesca_2001, Hamburger_2011, yadav2012, bayrak2013, Thapa_2018, Tourani_2023, mousavi2025b, mousavi2026d}.

Despite this, the spatio-temporal evolution of these seismicity parameters throughout the foreshock--aftershock sequence of the 2016 Kumamoto earthquake has not been comprehensively characterized. Prior studies of the sequence have focused primarily on coseismic slip inversion, teleseismic source properties, postseismic deformation, and strong-motion characteristics \cite<e.g., >{Kato2016, Asano_2016, Uchide_2016, lin2016coseismic, Yagi_2016, Fukahata_2016, Toda_2016, Yue_2017, Hao_2017, Jiang_2017, Kobayashi_2017, Zhang_2018, Hashimoto_2020}, while the statistical behavior of $b$, $D_c$, and $Z$ has, to date, not been jointly examined for this sequence at high spatio-temporal resolution.

To address this gap, this study revisits the 2016 Kumamoto earthquake sequence through a high-resolution, integrated spatio-temporal analysis of its foreshocks and aftershocks. By jointly examining the temporal and spatial evolution of the $b$-value, $D_c$-value, and $Z$-value, this study aims to provide new constraints on the seismogenesis, stress evolution, and seismotectonic behavior of the Kumamoto sequence and the surrounding Futagawa--Hinagu fault system.

\begin{figure}[htbp]
	\centering
	\hspace{-2cm}
	\includegraphics[width=1.1\textwidth]{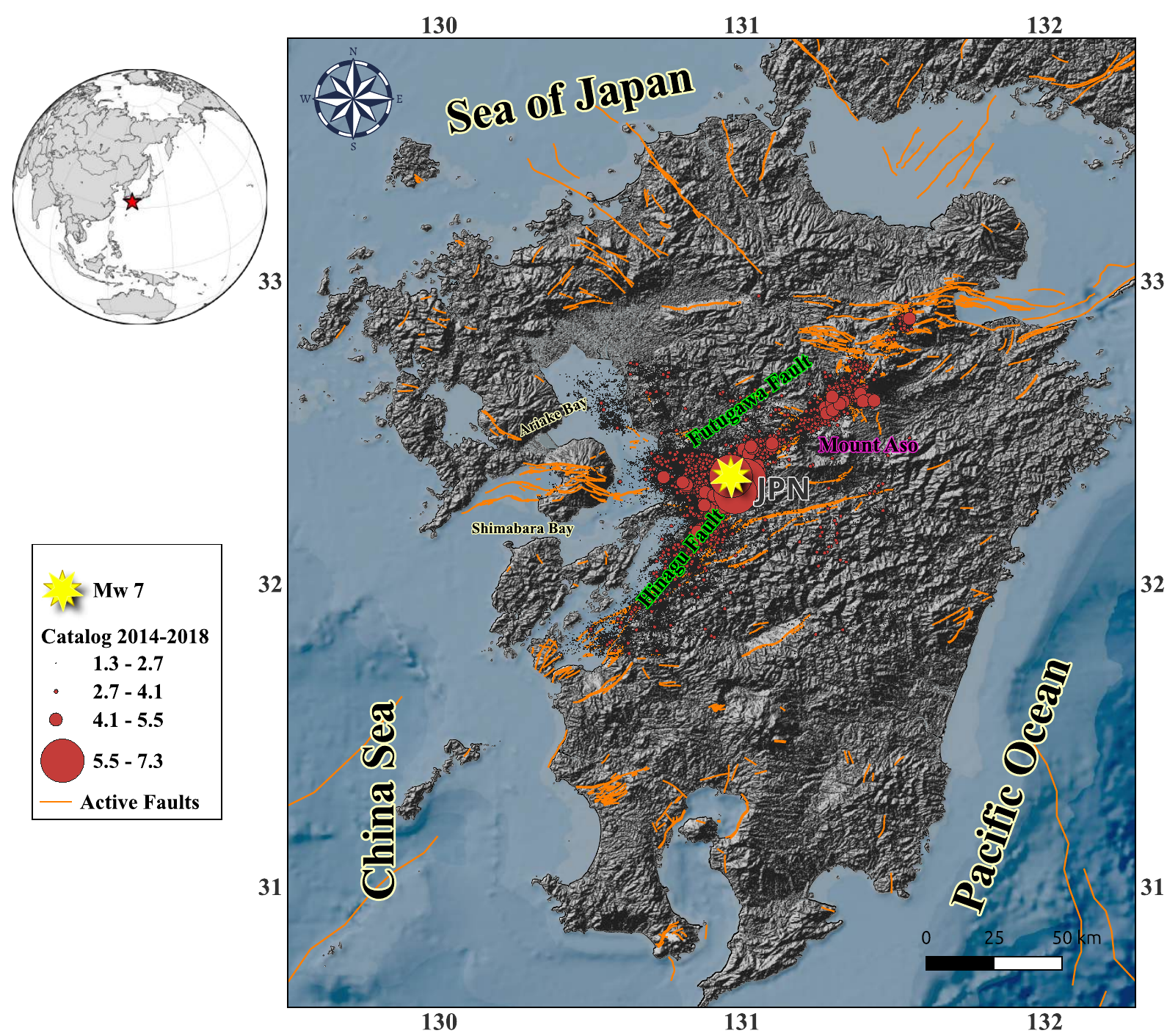}
	\caption{2D epicenter distribution of the Kumamoto earthquake sequence. The mainshock epicentral region is situated on the Futagawa and Hinagu fault zones. The inset in the top-right corner shows a regional map of Japan highlighting Kyushu Island.}
	\label{fig:2d_seismicity_inset}
\end{figure}

\section{Tectonic Setting}\label{sec:seismotectonics}
The Japanese archipelago represents one of the most seismically active regions globally, driven by the rapid convergence of several major tectonic plates, including the Pacific and Philippine Sea plates \citep{Kato2016}. While much of Northeast Japan is dominated by compressional stress and megathrust earthquakes along the subducting Pacific Plate (e.g., the 2011 T\=ohoku Oki earthquake), Southwest Japan presents a more complex tectonic configuration, forming the overriding Eurasian Plate above the northwestward subducting Philippine Sea Plate along the Nankai Trough and the Ryukyu Trench. The island of Kyushu, situated in Southwest Japan, exhibits a distinct tectonic regime in which the central part of the island is subjected to active North to South extensional stress \citep{Kato2016}. This extensional regime manifests as the Beppu Shimabara graben, which extends from Beppu Bay in the east to the Unzen Volcano on the Shimabara Peninsula in the west, spanning central Kyushu (Figure~\ref{fig:2d_seismicity_inset}). The normal fault systems developing within this graben have been attributed to two main mechanisms acting within the Eurasian Plate crust: the northward propagation of the opening Okinawa Trough back arc rift, and transient variations in the convergence direction of the subducting Philippine Sea Plate \citep{Kato2016}.

The 2016 Kumamoto earthquake sequence occurred along the Futagawa Hinagu fault zone, part of the western extension of the Median Tectonic Line, which is encompassed by the Beppu Shimabara graben in central Kyushu (Figure~\ref{fig:2d_seismicity_inset}) \citep{Kobayashi_2017}. The sequence was initiated by two right lateral foreshocks ($M_{\mathrm{JMA}}$ 6.5, corresponding to $M_w$ 6.2, and $M_{\mathrm{JMA}}$ 6.4, corresponding to $M_w$ 6.0) on 14 and 15 April 2016, which ruptured the northern segment of the Hinagu Fault on a subvertical plane dipping slightly to the east, distinct from the mainshock fault \citep{Kobayashi_2017}. The $M_w$ 7.0 (equivalently $M_{\mathrm{JMA}}$ 7.3) mainshock on 16 April initiated near the junction of the Hinagu and Futagawa faults and ruptured bilaterally along a steeply northwest dipping plane, with a dip of approximately 77 degrees. The dominant slip propagated about 30~km toward the northeast along the Futagawa Fault, while a smaller extension propagated about 10~km toward the southwest along the Hinagu Fault, together spanning a total rupture length of approximately 40~km \citep{Kubo_2016, moya2017, shirahama2016}. The rupture produced predominantly right lateral strike slip motion accompanied by a normal slip component, amounting to roughly 30 percent of the total moment, consistent with the observed subsidence of the northwestern block, which formed the hanging wall, and relative uplift of the southeastern block, which formed the footwall \citep{Kubo_2016, moya2017, shirahama2016}. The mainshock fault plane was oriented obliquely to the foreshock plane, reflecting rupture on a structurally distinct, non coplanar segment of the fault system. Rupture propagation terminated near Aso Caldera, coincident with a pronounced gap in aftershock activity attributed to elevated temperature and reduced seismic velocity associated with the underlying magmatic system \citep{Yagi_2016, shirahama2016}. The mainshock additionally triggered seismicity on structurally distinct, conjugate normal fault structures at remote sites within the Beppu Shimabara graben, for example the Yufuin and Kokonoe areas of Oita Prefecture, located approximately 60~km northeast of the epicentral region, consistent with static Coulomb stress transfer from the mainshock rupture \citep{toda2016science, Toda_2016}.

\section{Data and Selection Criteria}\label{sec:data}

\subsection{Seismic Observation Network and Catalog Provenance}\label{sec:dataset}
The raw seismicity records utilized in this retrospective analysis were sourced from the comprehensive and unified earthquake catalog compiled by the Japan Meteorological Agency (JMA). The JMA maintains a high-density, nationwide observation network that integrates a wide range of inland telemetric stations with advanced marine observatories. Notably, this framework incorporates real-time seafloor seismic arrays, including S-net (Seafloor Observation Network for Earthquakes and Tsunamis along the Japan Trench) and DONET2 (Dense Oceanfloor Network System for Earthquakes and Tsunamis). Although S-net and DONET2 are situated primarily offshore to monitor trench-parallel megathrust zones, their real-time integration significantly enhances the baseline detection threshold and hypocentral resolution of the entire Japanese archipelago. When combined with the high-density land-based telemetric networks, such as Hi-net, it provides exceptionally high-precision determinations of event origin times, three-dimensional spatial coordinates, and magnitude scales, establishing a reliable foundation for high-resolution spatio-temporal seismicity analyses of the Kumamoto source region.

\subsection{Spatial Aperture and Temporal Window Selection}\label{sec:selection}
\begin{figure}[htbp]
	\centering
	\includegraphics[width=1.1\textwidth]{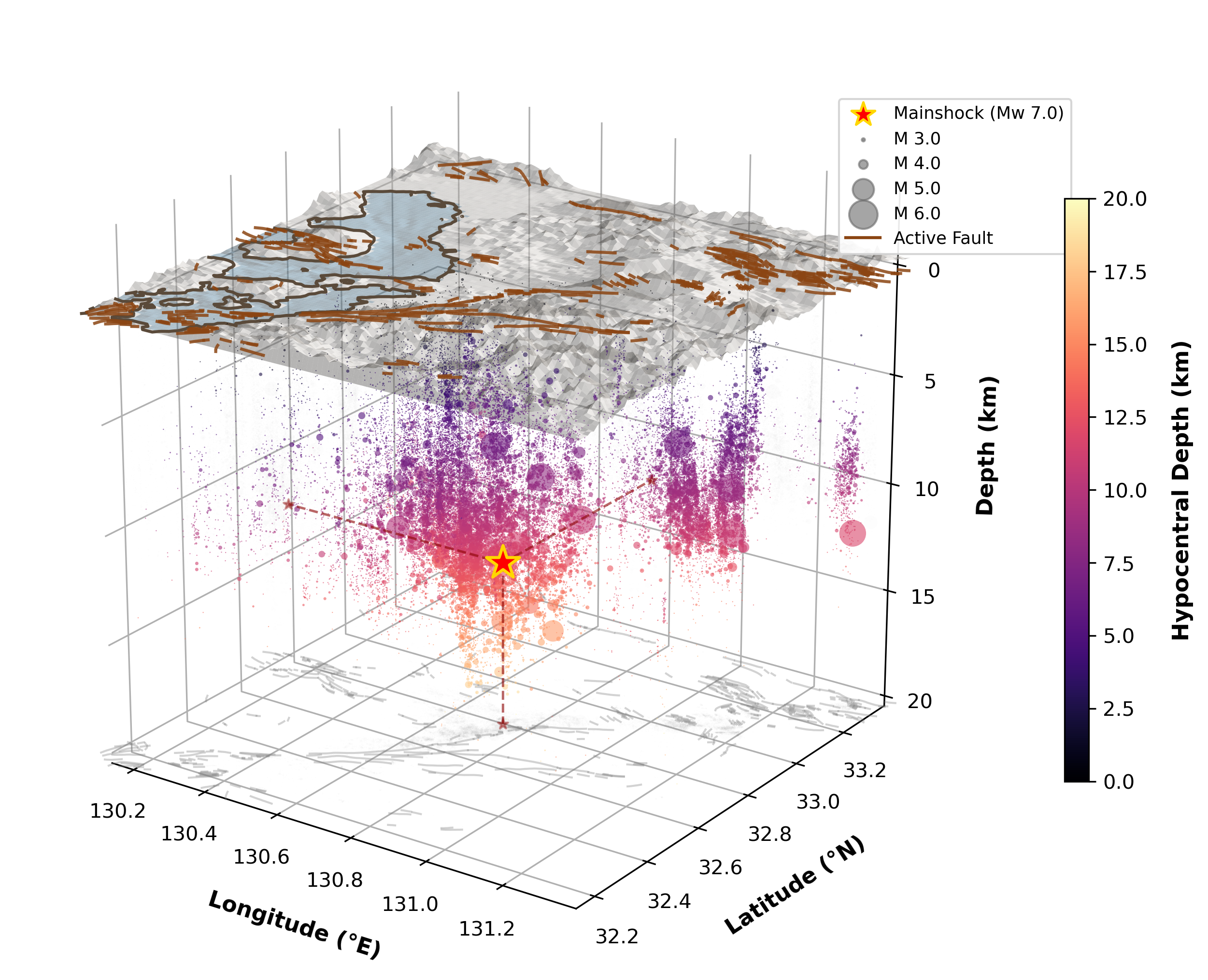}
	\caption{3D spatio-temporal distribution of the Kumamoto aftershock sequence ($M \ge 2.0$, depth $\le 20.0$ km). The mainshock is represented by the red star, with aftershocks scaled proportionally by magnitude and colored by hypocentral depth to illustrate structural layering. High-resolution digital elevation data (GMRT) is rendered at the surface, overlaid with active fault traces (brown lines) from RGAFJ (1991). The spatial distribution and fault traces are projected onto the bounding planes to resolve the geometric relationship between the active fault architecture and the aftershock cascade.}
	\label{fig:3d_seismicity}
\end{figure}

Isolating a geomechanically representative subset of seismicity from the regional catalog requires defining rigorous spatial and temporal boundaries that envelope the active deformation zone of the 2016 mainshock without introducing extraneous background signals. To achieve this, we constructed a localized spatial aperture spanning $1.2^\circ \times 1.2^\circ$ centered on the mainshock's epicentral coordinate ($Lat_{\text{event}}, Lon_{\text{event}}$) as determined by the JMA ($32.75^\circ\text{N}, 130.76^\circ\text{E}$). This translates to an active geographic window defined by $Lon \in [130.16^\circ, 131.36^\circ]$ and $Lat \in [32.15^\circ, 33.35^\circ]$. Geologically, this spatial window is optimally scaled: it is sufficiently expansive to capture the entire rupture plane of the Futagawa and Hinagu fault zones, secondary off-fault deformation strands, and the complete aftershock zone, yet narrow enough to filter out unrelated microseismicity hosted by adjacent, separate fault blocks or distant geothermal systems in the wider Beppu--Shimabara graben. Temporally, our analysis utilizes a four-year window centered on the mainshock ($Time_{\text{event}}$), extending from April 2014 (two years prior to the mainshock) to April 2018 (two years following the mainshock). This temporal frame establishes a statistically stable pre-seismic background baseline, allowing for a clear comparison between the long-term precursory stress loading anomalies and the subsequent post-seismic stress relaxation and aftershock decay sequence (the 3D hypocentral distribution of the post-seismic catalog is illustrated in Figure~\ref{fig:3d_seismicity}).

It is worth noting that while the seismicity catalog spans both the pre- and post-earthquake intervals, the spatial mapping of the seismicity parameters ($b$-value, $Z$-value, and $D_c$-value) in the main manuscript is restricted to the pre-seismic phase. This choice is physically motivated by the fact that post-seismic periods are dominated by massive aftershock sequences, which reflect transient, stress-relaxation processes governed by Omori-like decay \citep{omori1894, Utsu1995} rather than the tectonic loading anomalies that act as intermediate-term precursory indicators. In contrast, the temporal evolution of these parameters is calculated continuously across both the pre- and post-earthquake intervals. Including the post-earthquake period in the time-series is essential to demonstrate the recovery of the parameters and validate the complete precursory stress-loading--coseismic-rupture--postseismic-relaxation cycle, confirming that the observed pre-seismic anomalies were transient departures from the background baseline.

\section{Methodology}\label{sec:methodology}

\subsection{Analysis of $b$-value}\label{sec:b_value}
The Gutenberg Richter law is a well established statistical relationship that describes the frequency magnitude distribution of earthquakes in a given region \citep{gutenberg1944}. It is expressed as:
\begin{linenomath*}
	\begin{equation}
		\log_{10} N(\ge M) = a - b M
	\end{equation}
\end{linenomath*}
where $N$ is the cumulative number of earthquakes with magnitudes greater than or equal to $M$, the parameter $a$ reflects the overall seismic activity rate, and $b$ represents the slope of the distribution \citep{gutenberg1944}. The $b$-value is a fundamental parameter for characterizing regional seismic behavior \citep{gutenberg1944}. Although the average $b$-value at both global and regional scales is generally close to 1.0 \cite<e.g., >{frohlich1993, elisa2014}, its spatial and temporal variations are strongly influenced by physical factors such as tectonic stress, rock heterogeneity, and faulting style \cite<e.g., >{wyss1997, wiemer2000, schorlemmer2005, hussain2020, alkan2023, tourani2024}. Typically, high $b$-values ($b > 1.0$) indicate highly fractured or heterogeneous crust, commonly observed in volcanic or geothermal areas \cite<e.g., >{scholz1968, wiemer1998, farrell2009, enescu2011, tourani2024}. Conversely, low $b$-values ($b < 1.0$) are characteristic of regions with elevated stress concentration, such as active thrust fault zones \cite<e.g., >{wiemer2000, hussain2020, lewerissa2021, mousavi2026c}. For quantitative analysis, the $b$-value is usually estimated using the maximum likelihood method proposed by \citeA{aki1965}, which reduces statistical bias and provides robust results \citep{wiemer1997}. Accounting for the finite magnitude binning interval $\Delta M$ of the catalog, the estimator is given by:
\begin{linenomath*}
	\begin{equation}
		b = \frac{\log_{10} e}{M_{\mathrm{mean}} - \left(M_c - \dfrac{\Delta M}{2}\right)}
	\end{equation}
\end{linenomath*}
where $M_{\mathrm{mean}}$ is the average magnitude of events with $M \ge M_c$, $M_c$ is the magnitude of completeness, and $\Delta M = 0.1$ is the magnitude binning interval of the catalog. The spatial distribution of the $b$-value was estimated on a grid with spacing of $0.05^\circ$, with each node populated using the nearest events up to a minimum threshold of 50 earthquakes, subject to a maximum search radius of $100$~km to prevent inclusion of tectonically unrelated events at nodes near the edge of the study area. Rather than imposing a fixed regional threshold, the magnitude of completeness ($M_c$) was dynamically determined for each spatial grid cell and temporal window using the Maximum Curvature method \citep{wiemer2000}, which identifies $M_c$ directly from the peak of the non-cumulative frequency--magnitude distribution. The $b$-value was subsequently estimated using the maximum likelihood estimation method \citep{aki1965}. The associated uncertainty $\delta b$ was calculated following \citeA{shibolt1982}:
\begin{linenomath*}
	\begin{equation}
		\delta b = 2.30 b^2 \sqrt{ \frac{\sum_{i=1}^n (M_i - M_{\mathrm{mean}})^2}{n(n-1)} }
	\end{equation}
\end{linenomath*}
where $n$ is the number of earthquakes used in the calculation, $M_i$ represents individual magnitudes, and $M_{\mathrm{mean}}$ is their mean.

\subsection{Analysis of $D_c$-value}\label{sec:dc_value}
The spatial clustering of earthquakes can be quantified using the fractal dimension ($D_c$-value), which reflects the degree of self similarity in earthquake distributions. In this study, we applied the integral correlation method \citep{grassberger1983} to calculate the spatial fractal dimension, as this approach is particularly sensitive to variations in clustering behavior \citep{hirata1989a, hirata1989b, kagan1980}. The integral correlation function for the spatial distribution of $n$ earthquake hypocenters is defined as:
\begin{linenomath*}
	\begin{equation}
		C(r) = \frac{2}{n(n-1)} \sum_{i=1}^{n-1} \sum_{j=i+1}^n H(r - |X_i - X_j|)
	\end{equation}
\end{linenomath*}
where $C(r)$ is the correlation function, and $H$ is the Heaviside step function representing the number of earthquake pairs separated by a distance smaller than $r$. The parameter $r$ represents the three-dimensional Euclidean inter-event distance between two earthquake hypocenters, and $X_i$, $X_j$ denote the corresponding hypocentral coordinate vectors $(x, y, z)$. If earthquake hypocenters follow a fractal distribution, the correlation function scales as:
\begin{linenomath*}
	\begin{equation}
		\lim_{r\to 0} C(r) \propto r^{D_c}
	\end{equation}
\end{linenomath*}
where $D_c$ denotes the spatial fractal dimension. The fractal dimension is obtained from the slope of the best fit line in a log log plot of $C(r)$ versus $r$. Because $D_c$ is here computed from hypocentral, that is three-dimensional, coordinates including focal depth, its theoretical range is bounded between 0 and 3: values approaching 0 represent highly localized point-like clustering, values near 1.0 indicate a predominantly linear distribution along fault lines, values near 2.0 correspond to planar fault surfaces, and values approaching 3.0 reflect a volumetrically dispersed distribution of hypocenters across the seismogenic crust \cite<e.g., >{yadav2012a, yadav2012, ebrahimi2012, bayrak2013, ozturk2019, mousavi2026b, mousavi2026c, tourani2024}. These variations can be used to infer differences in fault complexity, segmentation, and the spatial distribution of deformation within active tectonic settings. For the analyses conducted in this study, the $D_c$-values were derived from the slope of the $\log C(r)$ versus $\log r$ relationship using a scaling range of 1 to 15 km, which yielded a well constrained linear regression following stability tests. The catalog used for the $D_c$ analysis was thresholded at the same magnitude of completeness $M_c$ dynamically evaluated in each window.

\subsection{Analysis of $Z$-value}\label{sec:z_value}
The temporal evolution of seismicity rates is evaluated using the $Z$-value method \citep{wiemer1994}, which quantifies statistically significant deviations from the long term background rate. For each spatial node, the $Z$-value at time $t$ is computed as:
\begin{linenomath*}
	\begin{equation}
		Z(t) = \frac{R_{\mathrm{wl}}(t) - R_{\mathrm{bg}}}{\sqrt{\frac{S_{\mathrm{wl}}^2(t)}{n_{\mathrm{wl}}} + \frac{S_{\mathrm{bg}}^2}{n_{\mathrm{bg}}}}}
	\end{equation}
\end{linenomath*}
where $R_{\mathrm{wl}}(t)$ and $R_{\mathrm{bg}}$ denote the mean seismicity rates within the moving window and the background period, respectively; $S_{\mathrm{wl}}(t)$ and $S_{\mathrm{bg}}$ are their standard deviations; and $n_{\mathrm{wl}}$ and $n_{\mathrm{bg}}$ are the corresponding numbers of temporal bins. Positive $Z$-values show statistically significant seismic quiescence, whereas negative values illustrate seismic activation relative to the long term average \citep{mousavi2026}. In order to address possible differences in the variations of the rates of seismicity at various locations, the analysis area is divided into cells with $0.05^\circ \times 0.05^\circ$ separation between centers, matching the grid used for the spatial $b$-value analysis so that all three parameter maps are directly comparable node for node. In each cell, the 50 nearest events in the catalog were used to provide a standardized procedure for calculating rates along the fault zone. By using the nearest neighbor technique, boundary effects are minimized and the analysis is performed in a homogeneous statistical fashion.
The background rate $R_{\mathrm{bg}}$ was determined from the approximately two year period preceding the onset of the foreshock sequence on 14 April 2016, drawn from a total analysis period spanning approximately from mid 2014 to mid 2018, that is roughly two years before and two years after the mainshock, while the foreground rate $R_{\mathrm{wl}}(t)$ is estimated in sliding windows spanning intervals of 1 month, 3 months, and 5 months, with the window advancing 1 month at a time to provide sufficient resolution and sample size for each step, applied across the full 2014 to 2018 analysis period.


\section{Results and Discussion}\label{sec:results}
The spatio-temporal evolution of seismicity parameters before and after major earthquakes provides one of the more direct observational windows into the mechanical state of the seismogenic crust. The 2016 Kumamoto earthquake sequence, culminating in the $M_w$ 7.0 mainshock on 16 April 2016, offers a well instrumented case study for examining this evolution across a complete earthquake cycle. Located within the tectonically active Beppu Shimabara graben system in central Kyushu, Japan, the region is characterized by active normal and strike slip faulting, elevated heat flow, and pronounced crustal heterogeneity. Its dense seismic monitoring network has produced a high quality earthquake catalog, enabling spatially and temporally resolved analyses of seismicity patterns \citep{Kato2016, yoshida2017}. The present study integrates three complementary seismicity parameters, the Gutenberg Richter $b$-value, the fractal dimension ($D_c$-value), and the seismic quiescence ($Z$-value), to examine the evolution of the seismogenic crust before and after the Kumamoto mainshock.

Each parameter provides distinct information about the stress field and seismogenic structure of the crust, and their joint interpretation offers more diagnostic power than any single metric considered alone \cite<e.g., >{ozturk2018, wyss1988, Enescu2001, mousavi2026d, mousavi2026e}. The following sections examine the spatial and temporal evolution of these parameters during the pre seismic and post seismic periods, with particular attention to how changes in $b$ and $D_c$ reflect the transition from stress accumulation and fault zone localization before the mainshock to stress redistribution and crustal recovery afterward \citep{stein1999, toda2011}.

\subsection{Frequency Magnitude Distribution and Catalog Completeness}\label{sec:fmd}
\begin{figure}[htbp]
	\centering
	\includegraphics[width=0.9\textwidth]{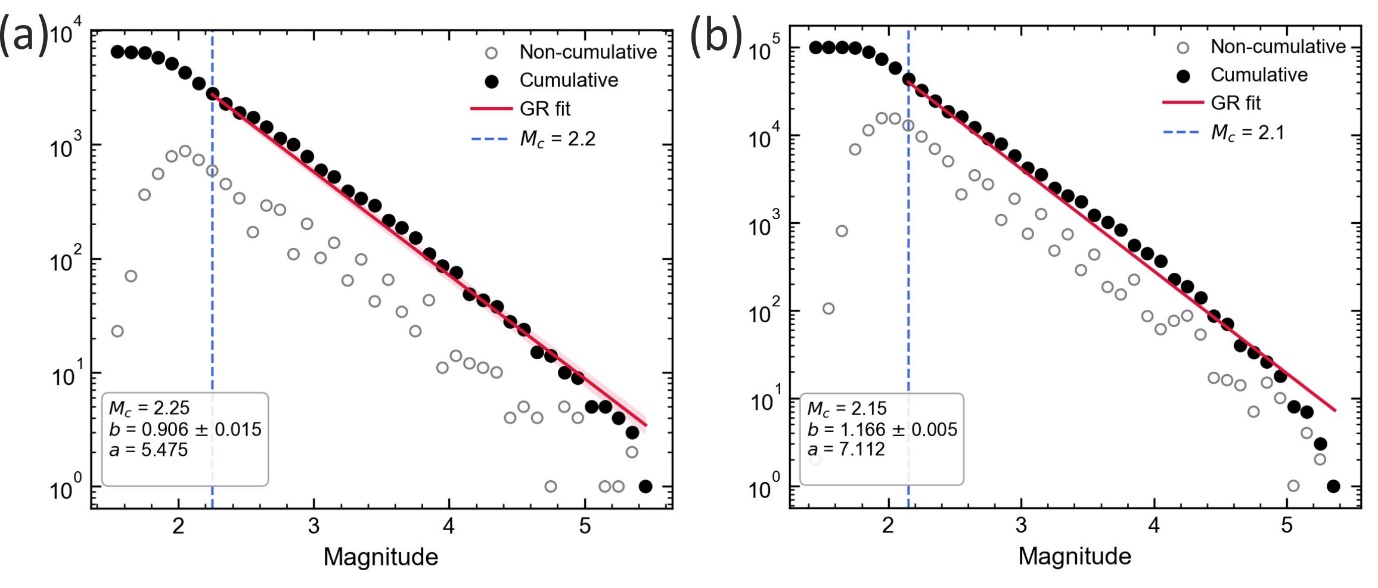}
	\caption{Frequency magnitude distribution of (a) the pre seismic and (b) the post seismic period for the Kumamoto source region. Black and white circles represent the cumulative and incremental event counts, respectively. The blue dashed line indicates the completeness magnitude, and the red line represents the Gutenberg Richter fit.}
	\label{fig:fmd}
\end{figure}

Before interpreting any derived seismicity parameter, the completeness and statistical reliability of the earthquake catalog must be assessed, since all subsequent analyses, including $b$-value estimation, fractal dimension calculation, and quiescence detection, are conditioned on the completeness threshold of the underlying data. Figure~\ref{fig:fmd} presents the frequency magnitude distributions for the pre mainshock and post mainshock catalogs, together with the Gutenberg Richter best fit line from maximum likelihood estimation. The analysis yielded a magnitude of completeness of $M_c = 2.25$ for the pre mainshock catalog and $M_c = 2.15$ for the post mainshock catalog. The slightly lower post mainshock $M_c$ suggests improved detection of smaller magnitude earthquakes during the dense aftershock sequence. The corresponding $b$-values were $0.906 \pm 0.015$ before the mainshock and $1.166 \pm 0.005$ after. Since the globally and regionally averaged $b$-value is commonly close to 1.0 \cite<e.g., >{frohlich1993, elisa2014}, the pre mainshock value below 1.0 indicates an elevated stress state and a relatively greater proportion of larger magnitude earthquakes \citep{wiemer1999, wiemer2002}, consistent with reductions in $b$-value reported before major earthquakes in other tectonic settings \cite<e.g., >{wiemer2000, murase2004, nakaya2006, tourani2024, mousavi2025a}. The higher post mainshock $b$-value reflects a greater relative abundance of smaller events, consistent with stress release and the dominance of aftershock activity following the mainshock.

\FloatBarrier

\subsection{Pre and Post Seismic Spatial Distribution of $b$-value}\label{sec:spatial_b}
\begin{figure}[htbp]
	\centering
	\includegraphics[width=0.95\textwidth]{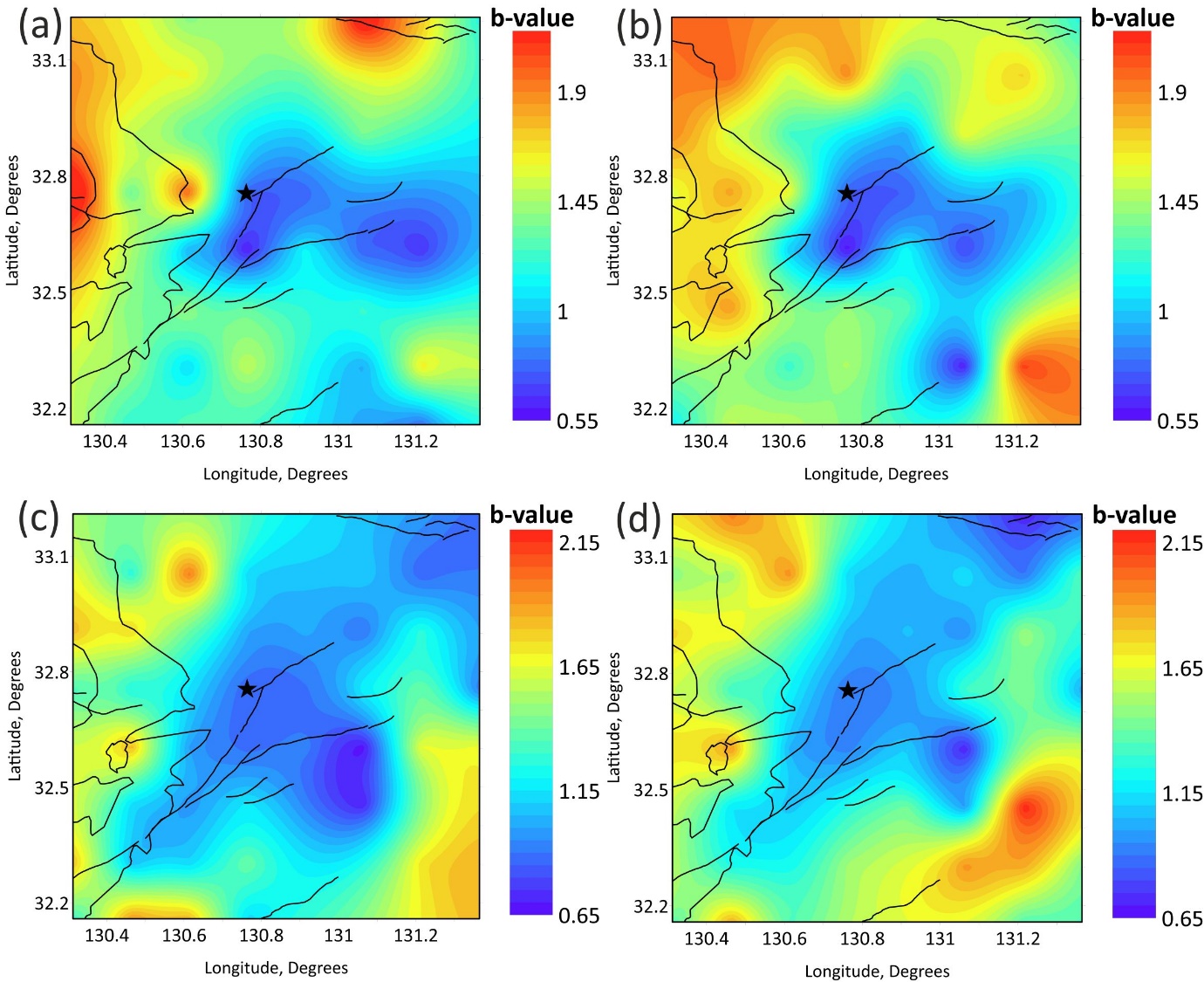}
	\caption{Spatial distribution of the $b$-value estimated over two successive temporal windows preceding the 2016 $M_w$ 7.0 Kumamoto mainshock: (a) one year before mainshock, (b) six months before mainshock, (c) six months after mainshock, and (d) one year after mainshock. The star denotes the mainshock epicenter.}
	\label{fig:spatial_b}
\end{figure}

The spatial distribution of the $b$-value estimated over two temporal windows preceding the Kumamoto mainshock, one year and six months (Figure~\ref{fig:spatial_b}a, b), shows a spatially coherent pattern that narrows and intensifies as the mainshock approaches. Across both windows, estimated $b$-values range from approximately 0.55 to 1.9. The lowest values ($b < 1.0$) are consistently concentrated around the mainshock epicenter, located at approximately $32.75^\circ\text{N}$, $130.76^\circ\text{E}$, and along the Futagawa Hinagu Fault Zone and its adjacent segments. This pattern is consistent with the inverse relationship between $b$-value and accumulated shear stress, in which elevated tectonic loading is expressed as a statistically significant $b$-value low \cite<e.g., >{schorlemmer2005, tormann2015, scholz2015, tourani2024}, and is broadly comparable to anticipatory $b$-value declines documented before other major earthquakes, including the 2011 Tohoku Oki, 2009 L'Aquila, and 2017 Sarpol e Zahab events \citep{nanjo2012, Gulia2019, mousavi2025a}. The consistent occurrence of the low $b$-value zone across both pre mainshock windows supports previous findings that large earthquakes preferentially occur in regions of relatively low $b$-value and that $b$-value commonly decreases through time as tectonic stress accumulates \citep{elisa2014, smith1981, rehman2015, chiba2019, wangr2021, jena2021,  tourani2024, Gulia2019, tormann2015, mousavi2025a}. The spatial coincidence between the low $b$-value anomaly and the inferred rupture zone suggests that $b$-value mapping delineates regions of elevated tectonic stress that subsequently experienced coseismic failure.

\FloatBarrier

The post seismic spatial distribution of $b$-value was analyzed for six months and one year after the mainshock (Figure~\ref{fig:spatial_b}c, d) and shows a clear change relative to the pre seismic pattern. Before the earthquake, $b$-values ranged from 0.55 to 1.90; after the mainshock they ranged from 0.65 to 2.15, indicating an overall increase consistent with the rapid stress drop and partial stress re equilibration expected after a large rupture \citep{scholz2015, Gulia2019}. The persistence of $b$-values below 1.0 in the immediate vicinity of the mainshock, however, indicates that the fault zone continued to exhibit relatively elevated stress conditions after the earthquake, plausibly related to the relatively high proportion of aftershocks with magnitude greater than $M_w$ 4.0 \citep{ozturk2025}. Six months after the earthquake (Figure~\ref{fig:spatial_b}c), the zones of elevated $b$-value expand and shift toward the northwestern part of the study area and toward the southeastern extension of the rupture zone, suggesting a redistribution of seismicity following the mainshock, possibly reflecting the progressive reloading of secondary fault structures by aseismic afterslip and viscoelastic relaxation in the lower crust and upper mantle \citep{freed2005, perfettini2004, king1994, toda2011}. This immediate post seismic $b$-value elevation is consistent with the reduction in shear stress on and around the ruptured interface expected after a large rupture: the coseismic stress drop, of order 1 to 10 MPa for an $M_w$ 7.0 event, lowers the effective confining stress in the aftershock zone, shifting the earthquake size distribution toward smaller events and thus increasing $b$ \citep{Wiemer2001, schorlemmer2005}.

One year after the earthquake (Figure~\ref{fig:spatial_b}d), $b$-values increased further around the mainshock fault and adjacent segments, consistent with a continued reduction in shear stress and post seismic relaxation, comparable to aftershock sequences documented after other large crustal earthquakes such as the 1999 Izmit and 2010 Canterbury events \citep{ozturk2018, Gulia2019}. The one year $b$-value map does not represent a simple return to pre seismic conditions but rather a reorganized stress field reflecting the superposition of tectonic loading, coseismic stress transfer, and post seismic relaxation.

\subsection{Pre- and Post-Seismic Temporal Evolution of Seismicity Parameters ($b$-value, $D_c$-value, and $Z$-value)}\label{sec:temporal}
\begin{figure}[htbp]
	\centering
	\includegraphics[width=0.98\textwidth]{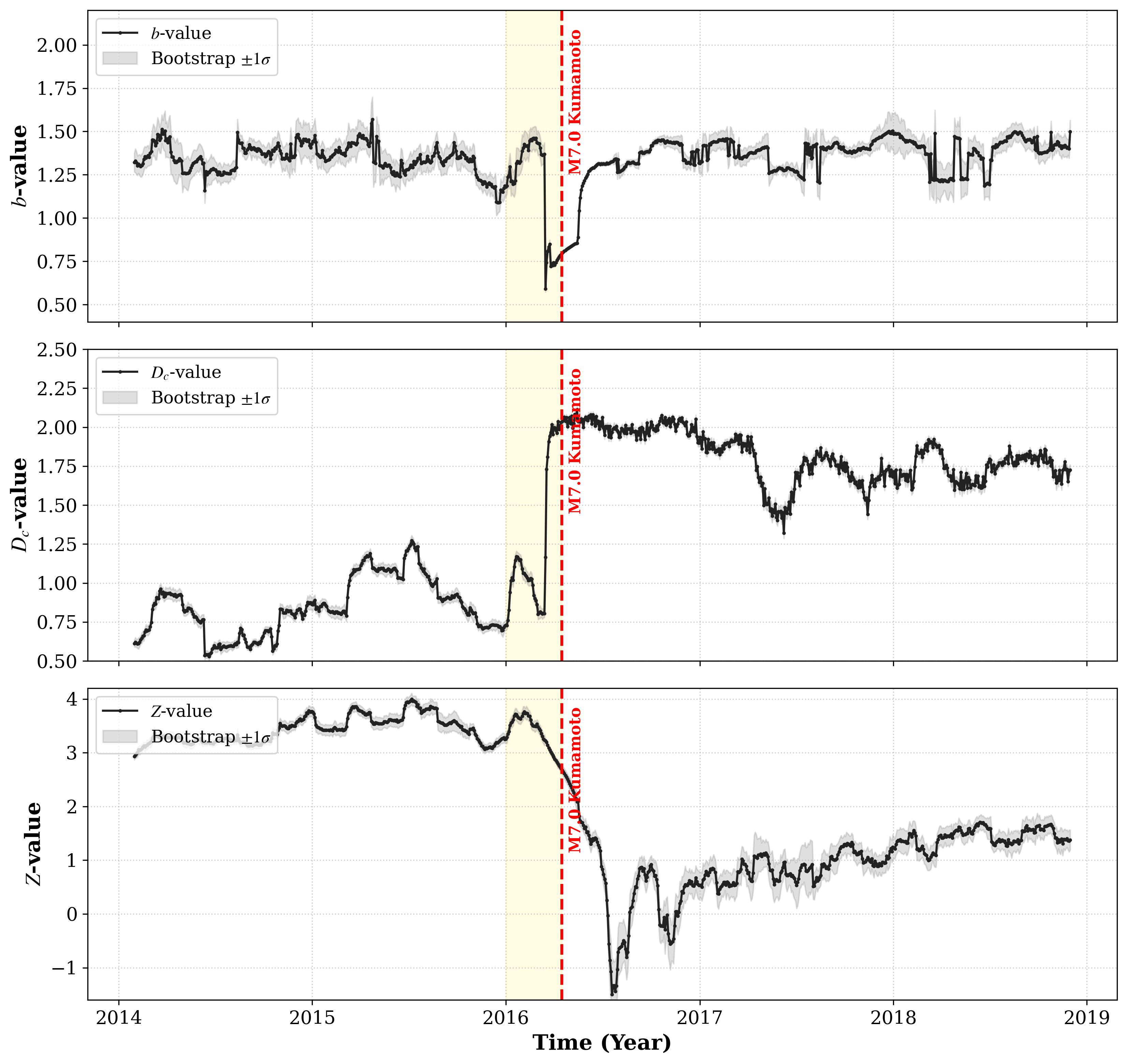}
	\caption{Temporal evolution of seismicity parameters for the 2016 Kumamoto earthquake sequence evaluated in a 2-month moving window with a 2-day sliding step: (a) Gutenberg--Richter $b$-value, (b) 3D hypocentral fractal dimension ($D_c$-value), and (c) standardized seismicity rate anomaly ($Z$-value). Shaded gray envelopes represent $\pm 1\sigma$ empirical uncertainties derived from non-parametric bootstrap resampling. The red dashed vertical line marks the 16 April 2016 $M_w 7.0$ Kumamoto mainshock, and the shaded yellow band highlights the ~3.5-month precursory window (1 January 2016 to 16 April 2016).}
	\label{fig:temporal_triad}
\end{figure}

To resolve the temporal dynamics of the seismogenic crust across the complete earthquake cycle of the 2016 $M_w$ 7.0 Kumamoto event, we continuously evaluate the multi-parameter triad of seismicity state variables:  $b$- $D_c$-, and  $Z$-value. Figure~\ref{fig:temporal_triad} presents the unified time series spanning 2014 to the end of 2018, calculated using a 2-month moving window with a 2-day sliding step across the active deformation polygon. All time series are accompanied by empirical $\pm 1\sigma$ uncertainty envelopes computed via non-parametric bootstrap resampling ($N_{\mathrm{boot}} = 50$ for $b$ and $Z$, $N_{\mathrm{boot}} = 30$ for $D_c$). The $Z$-value time series is evaluated against the pre-mainshock background baseline ($R_{\mathrm{bg}} = 12.94$ events/day, $S_{\mathrm{bg}} = 9.06$). The immediate precursory observation window (~3.5 months, from 1 January 2016 to the 16 April 2016 mainshock) is highlighted by the transparent yellow band.

During the pre-mainshock baseline period (2014 through late 2015), the $b$-value fluctuated around a relatively high mean of $b \approx 1.35 \pm 0.10$ (ranging between 1.16 and 1.57; Figure~\ref{fig:temporal_triad}a), reflecting typical background micro-seismicity distributed across the Beppu--Shimabara graben. As tectonic strain progressively localized along the Futagawa--Hinagu fault interface in early 2016, the $b$-value exhibited a pronounced, systematic decline throughout the precursory window, plunging from approximately 1.45 to an anticipatory precursory minimum of $b \approx 0.59$, and remaining depressed at $b \approx 0.73$--$0.79$ immediately prior to dynamic failure. In accordance with laboratory rock-friction experiments and observational seismology \citep{scholz2015, schorlemmer2005, amitrano2003, Gulia2019, nanjo2012}, this pre-mainshock drop in $b$ documents intense shear stress concentration on the locked fault interface leading up to nucleation. Coincident with the 16 April 2016 mainshock and the initial aftershock cascade, $b$ underwent an immediate coseismic drop followed by a rapid rebound to $b \approx 1.25$--$1.35$ within several weeks. This post-seismic surge is physically driven by the massive coseismic shear stress drop ($\Delta \tau \sim 1$--$10$~MPa) across the primary rupture zone \citep{kanamori1975, allmann2009, Asano_2016}, which reduced the confining differential stress and promoted widespread micro-fracturing dominated by small-magnitude aftershocks. Throughout the subsequent relaxation phase (2017--2018), the $b$-value remained elevated and stable ($b \approx 1.38 \pm 0.08$, range 1.19--1.50), consistent with long-term post-seismic crustal recovery.

The temporal evolution of the 3D hypocentral fractal dimension $D_c$ (Figure~\ref{fig:temporal_triad}b) provides crucial spatial context that complements the stress state inferred from the $b$-value. During the pre-mainshock baseline period, $D_c$ exhibited moderate values averaging $D_c \approx 0.85 \pm 0.15$ (ranging between 0.53 and 1.27), indicating that background hypocenters were distributed predominantly along quasi-linear to planar fault strands with episodic localized clustering. In the immediate precursory interval preceding the mainshock, $D_c$ dropped to $D_c \approx 0.70$--$0.80$, reflecting intensified spatial clustering and hypocentral localization within a restricted seismogenic volume surrounding the future nucleation zone. Immediately upon the occurrence of the $M_w$ 6.2 foreshock and the $M_w$ 7.0 mainshock, $D_c$ surged dramatically to $D_c \approx 2.04$--$2.11$. This high post-seismic fractal dimension ($D_c > 2.0$) indicates that aftershocks were not confined to a single linear fault trace but were distributed volumetrically across the complex, conjugate, and multi-branched crustal fault system of the Futagawa and Hinagu segments. During 2017 and 2018, $D_c$ gradually relaxed toward $D_c \approx 1.60$--$1.85$ (mean $1.74$), punctuated by transient drops ($D_c \approx 1.32$) during localized secondary aftershock sequences.

The standardized seismicity rate anomaly ($Z$-value, Figure~\ref{fig:temporal_triad}c) reveals a clear signature of seismic quiescence and subsequent activation across the earthquake cycle. Throughout 2014 and 2015, $Z$-value remained elevated ($Z \approx 3.00$--$4.00$, mean $3.43$), indicating a sustained relative deficit in earthquake production (seismic quiescence) within the future rupture volume compared to the long-term regional baseline. During the early 2016 precursory phase, $Z$ peaked near $3.76$ before sharply declining toward $2.70$ right around the foreshock onset, capturing the transition from locked precursory quiescence to immediate foreshock acceleration. Following the mainshock, the intense aftershock cascade drove $Z$ down to its minimum floor of $-1.50$ in mid-2016, quantitatively marking peak seismic activation. Over the subsequent 2.5 years (late 2016 to 2018), $Z$ exhibited a continuous, monotonic upward trajectory, rising from $-1.50$ to approximately $+1.14$ (range $0.38$ to $1.71$). This steady recovery directly tracks the modified Omori-law decay of aftershock productivity \citep{omori1894, Utsu1995}, confirming the progressive return of the seismogenic fault system toward background tectonic equilibrium.

\FloatBarrier

\subsection{Pre- and Post 3D spatial Evolution of the $b$-value}\label{sec:3d_b_value}
\begin{figure}[htbp]
	\centering
	\begin{subfigure}[b]{0.99\textwidth}
		\centering
		\includegraphics[width=\textwidth]{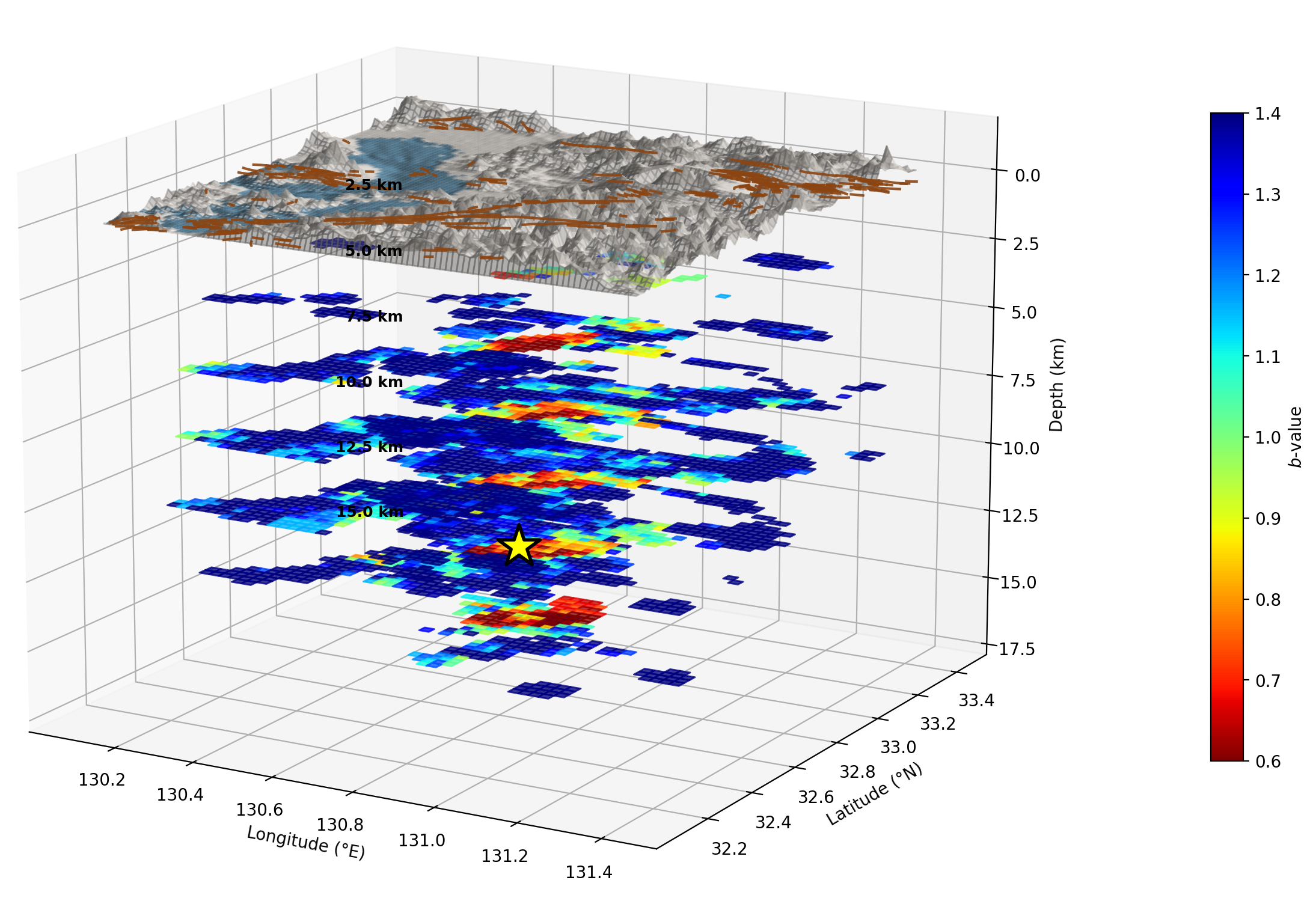}
		\caption{Pre-mainshock stress accumulation period}
		\label{fig:3d_b_pre}
	\end{subfigure}
	\par\vspace{0.2cm}
	\begin{subfigure}[b]{0.99\textwidth}
		\centering
		\includegraphics[width=\textwidth]{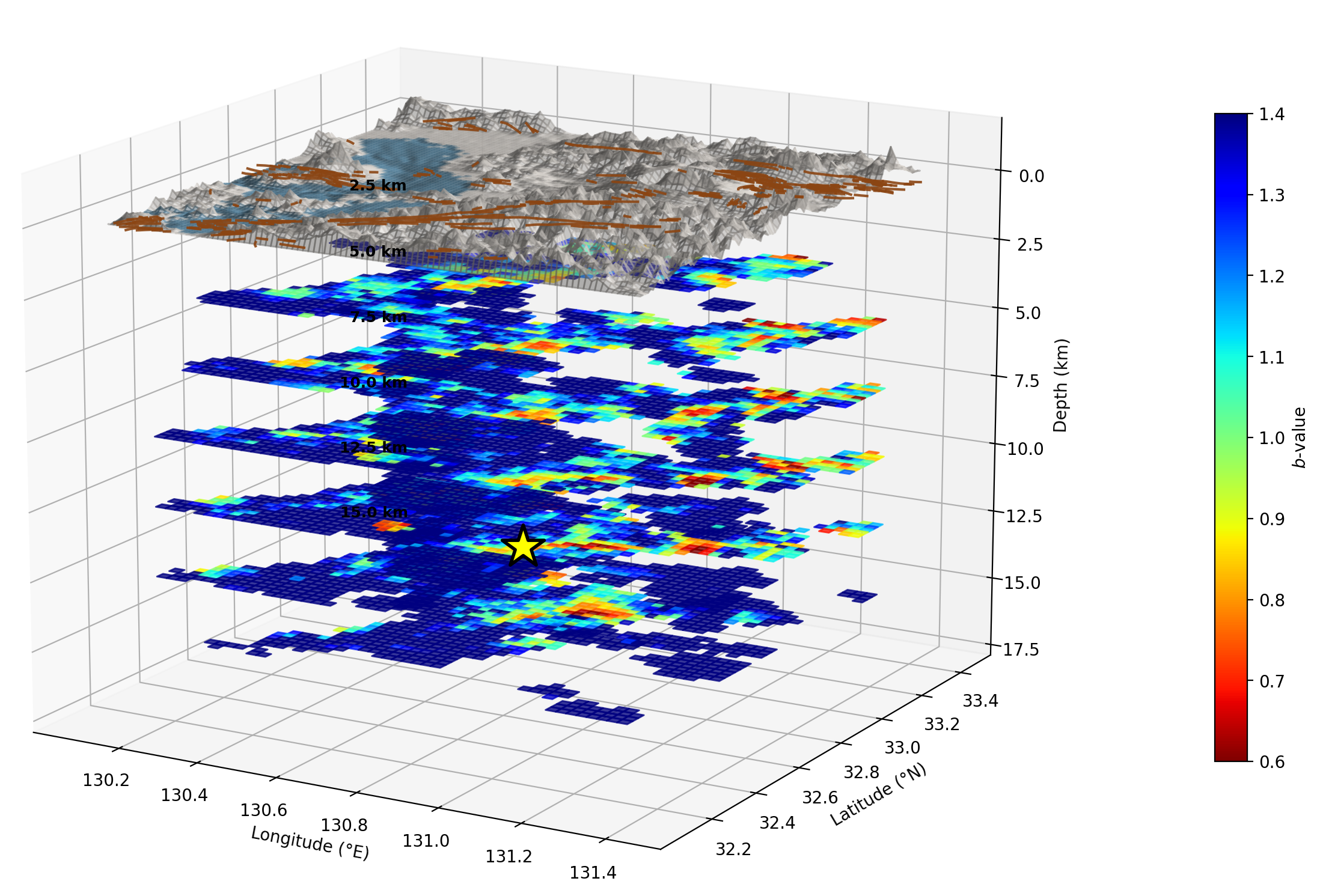}
		\caption{Post-mainshock stress relaxation period}
		\label{fig:3d_b_post}
	\end{subfigure}
	\caption{Three-dimensional depth-sliced distribution of the $b$-value evaluated across seven horizontal crustal layers ($z = 2.5$, $5.0$, $7.5$, $10.0$, $12.5$, $15.0$, and $17.5$~km; sampled within a $\pm 2.5$~km vertical depth interval), overlaid with the surface digital elevation model and active fault traces (brown lines) for (a) the pre-mainshock stress accumulation phase and (b) the post-mainshock relaxation phase. The yellow star denotes the 2016 $M_w$ 7.0 Kumamoto mainshock hypocenter at $z \approx 12.5$~km.}
	\label{fig:3d_b_combined}
\end{figure}

To overcome the spatial averaging inherent in two-dimensional map projections and one-dimensional transects, we reconstruct the full three-dimensional volumetric distribution of the Gutenberg--Richter $b$-value across the seismogenic crust. Figure~\ref{fig:3d_b_combined} illustrates the pseudo-tomographic depth slices of $b$-value evaluated across seven discrete crustal levels ($z = 2.5$, $5.0$, $7.5$, $10.0$, $12.5$, $15.0$, and $17.5$~km) beneath the surface digital elevation model (DEM) and mapped active fault traces, comparing the pre-seismic stress accumulation phase (Figure~\ref{fig:3d_b_combined}a) with the post-seismic relaxation period (Figure~\ref{fig:3d_b_combined}b). For each depth slice, the earthquake catalog is sampled within a vertical selection window of $\pm 2.5$~km centered at the target slice depth (spanning a total vertical thickness of $5.0$~km). The $b$-value is mapped to $b \in [0.60, 1.40]$ for visual purposes, providing a high-resolution window into the depth-dependent stress state of the Futagawa--Hinagu fault zone. Areas appearing hollow in the cross-sections correspond to spatial bins with an insufficient data threshold ($N < 15$ events), which are omitted from the visualization.

In the pre-mainshock volume (Figure~\ref{fig:3d_b_combined}a), a pronounced vertical and lateral differentiation in the $b$-value field is observed. In the shallowest crust ($z = 2.5$~km), seismicity is sparse and characterized by elevated $b$-values ($b \approx 1.20$--$1.40$, indicated by deep blue hues), indicative of low lithostatic confining pressure, uncoupled shallow fault segments, and pervasive micro-scale fracture networks. In stark contrast, the pre-mainshock section clearly exhibits a prominent, continuous low $b$-value anomaly ($b \approx 0.60$--$0.85$, highlighted in red-to-orange tones) that demonstrates pronounced stress accumulation across the depth range of 5.0 to 15.0~km along the Futagawa--Hinagu fault interface. The lowest $b$-values ($b \le 0.65$) are tightly focused at focal depths of $10.0$ and $12.5$~km, directly enveloping the future nucleation hypocenter of the $M_w$ 7.0 Kumamoto mainshock (indicated by the yellow star at $z \approx 12.5$~km, $32.75^\circ\text{N}$, $130.76^\circ\text{E}$). In accordance with the inverse scaling between accumulated tectonic stress and $b$-value demonstrated in laboratory rock mechanics and observational seismology \citep{scholz2015, schorlemmer2005, amitrano2003, goebel2013acoustic}, this deep-seated low-$b$ anomaly delineates a highly stressed, strongly locked asperity patch where tectonic strain was intensely concentrated prior to dynamic rupture. The lateral bounding of this low-$b$ core by higher background $b$-values ($b > 1.20$) in the surrounding crust further highlights the pronounced localized tectonic loading sustained on the primary fault plane.

Following the mainshock, the three-dimensional $b$-value volume (Figure~\ref{fig:3d_b_combined}b) exhibits a profound structural and mechanical transformation. Driven by the prolific aftershock cascade, the spatial sampling of the seismogenic volume expands across all depth levels from $5.0$ to $17.5$~km. Most notably, the pre-seismic low-$b$ core at $z = 10.0$--$12.5$~km undergoes a dramatic stress-inversion, transitioning to predominant high $b$-values ($b \approx 1.25$--$1.40$, dark blue). This widespread post-seismic $b$-value surge is physically attributable to the massive coseismic shear stress drop ($\Delta \tau \approx 1$--$10$~MPa) across the ruptured fault plane \citep{kanamori1975, allmann2009, Asano_2016}, which diminished effective confining stress and triggered a high density of micro-fracturing and secondary fault readjustments dominated by smaller-magnitude aftershocks \citep{Gulia2019, nanjo2012}. 

Nevertheless, the post-mainshock 3D distribution reveals critical localized stress heterogeneities that depart from total relaxation. In particular, distinct patches of moderate-to-low $b$-values ($b \approx 0.80$--$0.95$, visible in yellow-to-orange shades) persist at depth: (1) along the unruptured southwestern termination of the Hinagu fault segment at $z = 10.0$--$17.5$~km, and (2) near the northeastern rupture limit adjacent to the volcanic conduit system of Mount Aso at $z = 7.5$--$12.5$~km. These persistent low-$b$-value patches correspond to regions of positive Coulomb stress transfer ($\Delta \mathrm{CFS} > 0$) and stress concentration at geometrical fault barriers \citep{king1994, stein1999, toda2016science}, identifying unrelaxed asperities and secondary fault segments that maintained elevated stress conditions during the post-seismic sequence.

\FloatBarrier

\subsection{Spatio-temporal Evolution of $b$-value in Four-Month Consecutive Windows}\label{sec:3d_stacked_b}
\begin{figure}[htbp]
	\centering
	\includegraphics[width=0.99\textwidth]{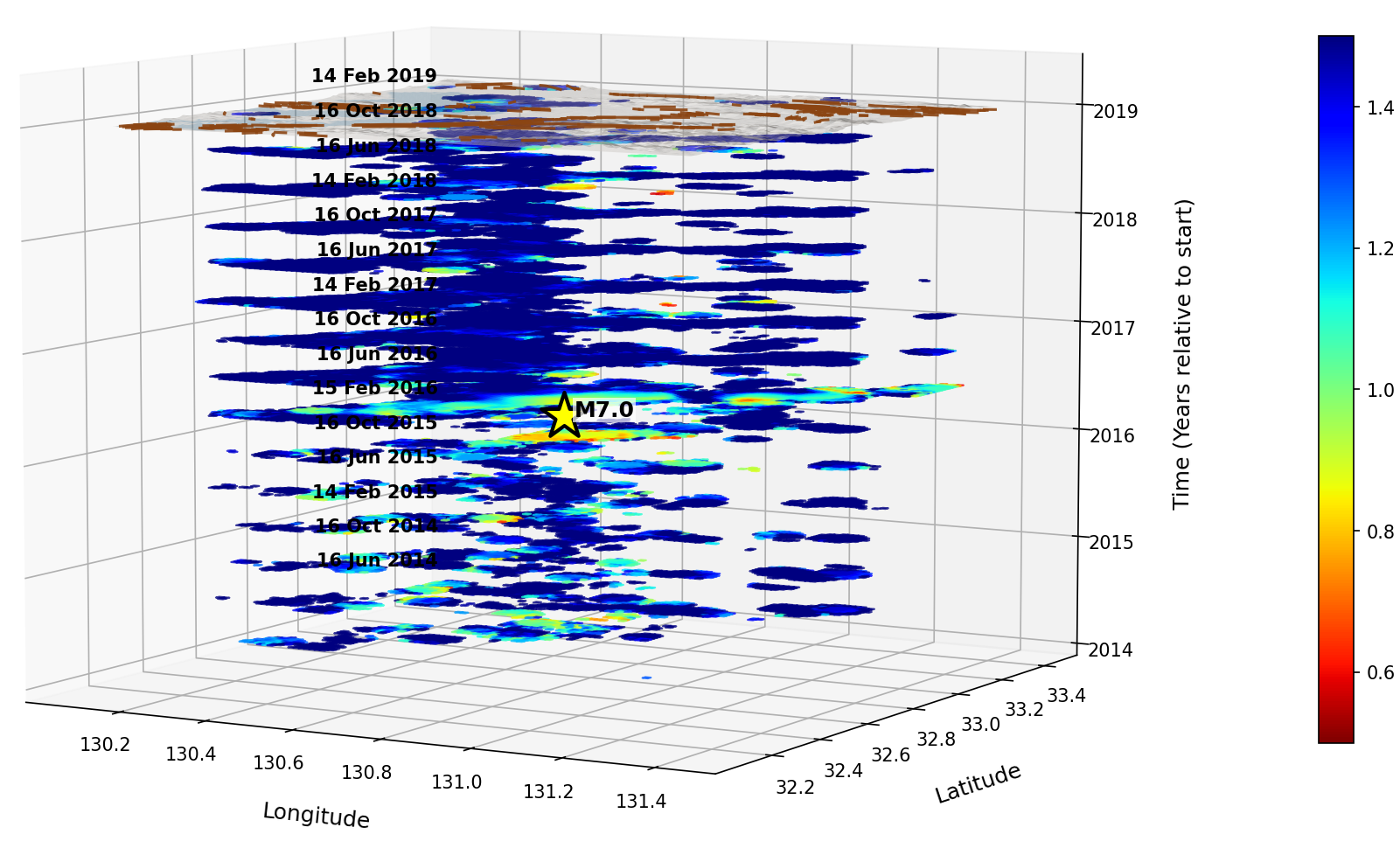}
	\caption{Three-dimensional spatio-temporal stacking of the $b$-value evaluated in consecutive four-month time windows from 2014 to 2019. The vertical axis represents time in years, with horizontal slices illustrating the spatial distribution of the $b$-value across central Kyushu beneath the digital elevation model and mapped active fault traces (brown lines). The yellow star marks the occurrence of the 16 April 2016 $M_w$ 7.0 Kumamoto mainshock on the time axis at the transition between the pre- and post-seismic observation windows.}
	\label{fig:3d_stacked_b}
\end{figure}

To decipher the fine-scale temporal onset and progression of the precursory stress accumulation imaged in the three-dimensional depth volumes, we construct a continuous spatio-temporal 3D stack of the Gutenberg--Richter $b$-value distribution evaluated in consecutive four-month non-overlapping observation windows spanning from 2014 to 2019 (Figure~\ref{fig:3d_stacked_b}). In this volumetric representation, the horizontal plane maps the spatial distribution of $b$-values across the seismogenic crust beneath the digital elevation model and mapped fault traces, while the vertical axis tracks continuous time (years relative to the 2014 origin). The 16 April 2016 $M_w$ 7.0 Kumamoto mainshock is denoted by the prominent yellow star located on the time axis at the precise temporal boundary between the 15 February 2016 pre-seismic window and the 16 June 2016 post-seismic window.

The four-month stacked sequence reveals that the pre-seismic crust maintained a stable, high background $b$-value regime ($b \approx 1.20$--$1.40$, indicated by uniform dark blue hues) throughout 2014 and most of 2015 (slices 16 Jun 2014, 16 Oct 2014, 14 Feb 2015, 16 Jun 2015, and 16 Oct 2015). During this prolonged multi-year baseline phase, localized low-$b$ anomalies were absent, indicating that tectonic loading was broadly distributed across the regional fault network without localized stress concentrations.

A striking, localized mechanical transition occurs exclusively in the four-month window immediately preceding the mainshock (the 15 February 2016 time slice, spanning mid-February to mid-April 2016). In this pre-mainshock slice, a sharply defined, low $b$-value anomaly ($b \approx 0.60$--$0.85$, highlighted by vibrant yellow, orange, and red contours) develops directly beneath the epicentral region of the impending rupture along the Futagawa--Hinagu fault zone. This observation establishes that the deep-seated low-$b$ zone identified in the 3D depth-sliced volume (Figure~\ref{fig:3d_b_combined}a) was not a static, long-term tectonic signature, but rather an acute, rapid stress accumulation process that concentrated on the locked seismogenic asperity predominantly during the final four months leading to catastrophic failure.

Following the mainshock, the 16 June 2016 time slice documents an immediate, widespread rebound to elevated $b$-values ($b \ge 1.25$--$1.40$, dark blue) across the entire rupture zone. This instantaneous $b$-value surge reflects the massive coseismic release of accumulated shear stress and the predominance of smaller-magnitude aftershocks during the initial relaxation phase. Over subsequent four-month intervals through 2017, 2018, and early 2019 (up to the 14 February 2019 slice), high $b$-values remain dominant across the fault system, recording the progressive viscoelastic relaxation and ongoing mechanical recovery of the ruptured crust.

\FloatBarrier

\subsection{Seismic Quiescence Analysis: Spatial Distribution of the $Z$-value}\label{sec:quiescence}
\begin{figure}[htbp]
	\centering
	\includegraphics[width=0.95\textwidth]{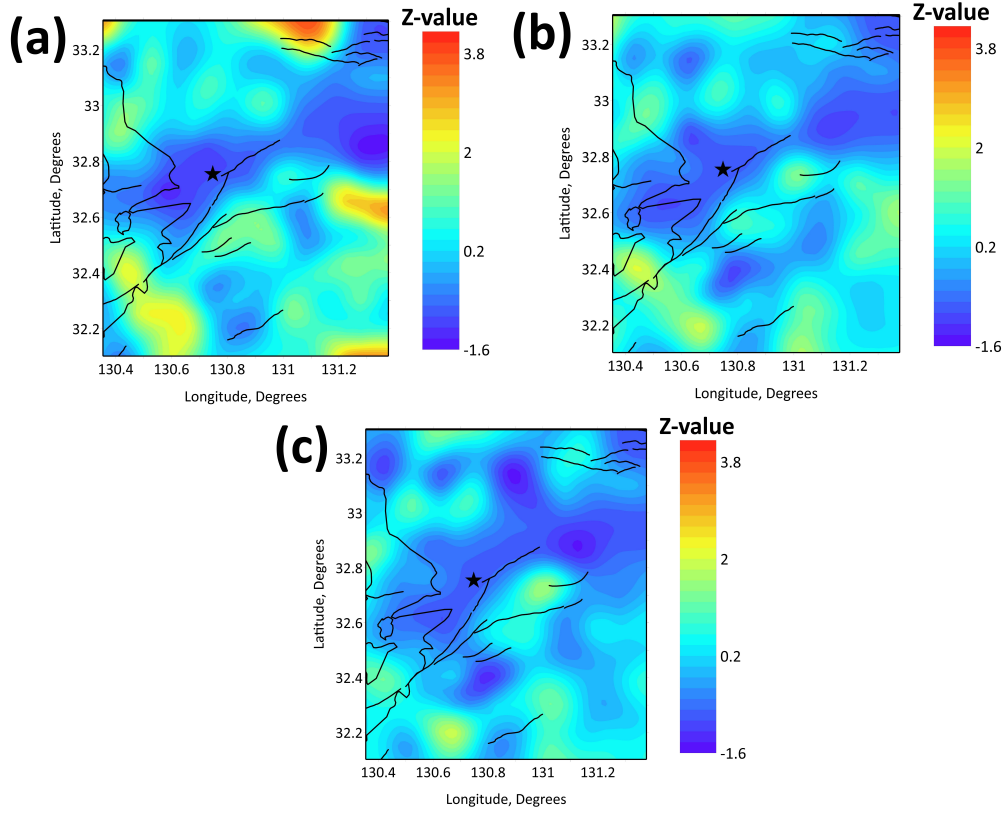}
	\caption{Spatial maps of the $Z$-value computed over a background period spanning approximately the two years prior to the mainshock, using three different test window lengths: (a) $R_{\mathrm{wl}} = 1$ month, (b) $R_{\mathrm{wl}} = 3$ months, and (c) $R_{\mathrm{wl}} = 5$ months. Negative values indicate quiescence and positive values indicate activation. The star denotes the mainshock epicenter.}
	\label{fig:spatial_z}
\end{figure}

Seismic quiescence, the anomalous reduction in seismicity rate relative to a background level, is a widely documented, though not universally reliable, precursory phenomenon associated with large earthquakes \citep{wyss1988, Enescu2001}. Quiescence is quantified here using the $Z$-value statistic, which measures the standardized departure of the observed seismicity rate in a test window from the mean rate computed over a longer background period. Given the sign convention of the $Z$-value formula used in this study, negative $Z$-values indicate seismic quiescence (a test window rate below the background rate), while positive $Z$-values indicate seismic activation (a test window rate above the background rate); the null hypothesis of stationarity is rejected at $|Z| > 2.0$. Figure~\ref{fig:spatial_z} presents spatial maps of $Z$ computed over a background period spanning approximately the two years preceding the mainshock, using three test window lengths: $R_{\mathrm{wl}} = 1$ month (Figure~\ref{fig:spatial_z}a), 3 months (Figure~\ref{fig:spatial_z}b), and 5 months (Figure~\ref{fig:spatial_z}c). This multi window approach is motivated by the fact that quiescence anomalies can be transient and scale dependent, such that a single test window length may miss anomalies whose temporal persistence lies outside that interval \citep{wyss1988, huang2006}.

The one month map (Figure~\ref{fig:spatial_z}a) shows a heterogeneous spatial pattern, with isolated patches of both positive $Z$, indicating locally elevated seismicity rate, possibly aftershock contamination or triggered seismicity, and negative $Z$, indicating quiescence, distributed irregularly across the study area. The most prominent quiescence anomaly in this window lies east of the mainshock epicenter, where $Z$ reaches approximately $-1.6$; this falls short of the $|Z| > 2.0$ significance threshold adopted here and should be read as suggestive rather than statistically significant, consistent with the greater susceptibility of short test windows to sampling noise and to incomplete separation of foreshock or triggered seismicity from the background rate. The three month window (Figure~\ref{fig:spatial_z}b) yields a spatially more coherent pattern: the negative anomaly becomes more defined and spatially extensive, centered approximately on $131.0^\circ\text{E}$, $32.7^\circ\text{N}$, with values again reaching approximately $-1.6$ over much of the anomalous region. As with the one month window, this remains below the nominal $|Z| > 2.0$ threshold; its spatial coherence and persistence across an independent, longer window nonetheless make it a candidate anomaly worth reporting, and it is interpreted with corresponding caution rather than as a confirmed statistically significant precursor. Physically, medium term quiescence of this kind is often interpreted as reflecting progressive interseismic locking of the fault interface, which can suppress small magnitude seismicity near the locked patch through elevated confining stress \citep{Enescu2001, ozturk2018}; the offset between the quiescence centroid and the eventual mainshock location is consistent with prior observations that fault locking can suppress seismicity at some distance from the locked zone rather than at its center \citep{wyss1988}.

The five month window (Figure~\ref{fig:spatial_z}c) provides the most time integrated view in this analysis and shows a spatially extensive negative $Z$ anomaly spanning a substantial part of the northeastern sector of the study area, with the most pronounced values, approximately $-1.6$ to $-2.0$, concentrated in a roughly elliptical zone elongated along the northeast to southwest direction, broadly consistent with the regional fault architecture of the Beppu Shimabara graben system; this window is the only one in which values approach or reach the adopted significance threshold. The same map also shows an emergent zone of elevated positive $Z$ near the northern margin of the study area, potentially indicating triggered seismicity or stress induced rate increases on structures peripheral to the main locking zone, broadly consistent with the stress shadow and stress loading pattern described by rate and state friction models, in which fault loading suppresses seismicity on a locking patch while accelerating it on structures experiencing increasing Coulomb stress \citep{dieterich1994, king1994, toda2011, toda2016science}. Taken together, all three test window lengths point to a spatially consistent zone of reduced seismicity rate in the general vicinity of the future mainshock; given that only the five month window clearly reaches the adopted significance threshold, this is best interpreted as a suggestive, spatially coherent quiescence signal that strengthens with integration time, rather than as an unambiguously significant precursor established independently at each window length.

\FloatBarrier

\subsection{Pre and Post Seismic Spatial Distribution of $D_c$-value}\label{sec:spatial_dc}
The fractal dimension ($D_c$-value), which quantifies the spatial clustering geometry of seismicity, provides a complementary perspective on the evolving seismogenic state of the crust. Unlike the $b$-value, which encodes the magnitude scaling of seismicity, $D_c$ is sensitive to the spatial self organization of earthquake hypocenters and serves as a proxy for the structural complexity of the active fault zone \citep{oncel2002, ozturk2018}. Because $D_c$ here is computed from hypocentral, that is three dimensional, inter event distances, its interpretive range spans 0 to 3: values near 1.0 indicate a predominantly linear distribution associated with fault structures, values approaching 0 represent highly localized clustering, values near 2 correspond to a planar distribution, and values approaching 3 reflect a more volumetrically dispersed distribution of hypocenters \citep{yadav2012, ebrahimi2012, bayrak2013, ozturk2019, tourani2024}.

\begin{figure}[htbp]
	\centering
	\includegraphics[width=0.95\textwidth]{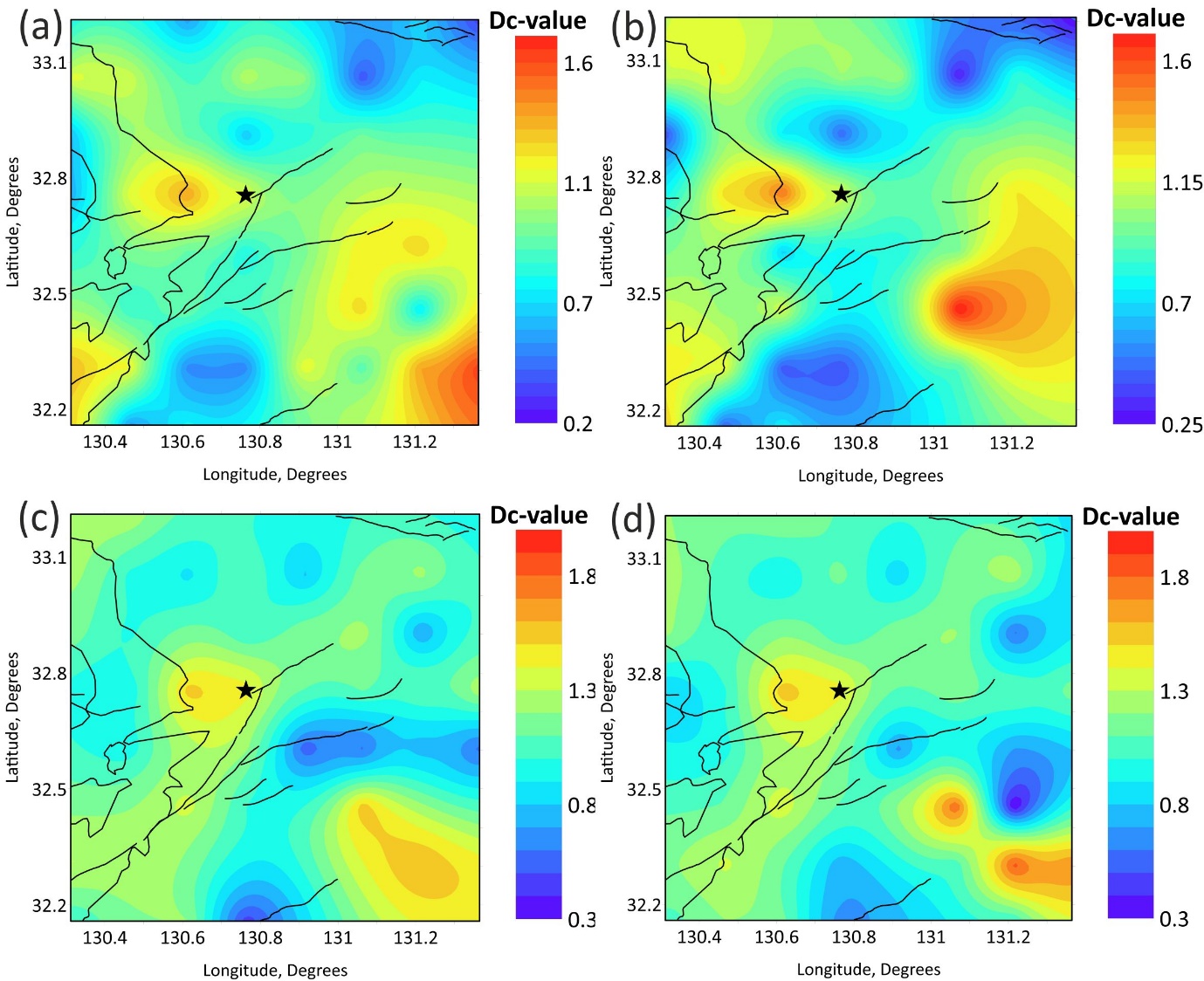}
	\caption{Spatial distribution of the fractal dimension $D_c$-value estimated over two successive temporal windows preceding the 2016 $M_w$ 7.0 Kumamoto mainshock: (a) one year before mainshock, (b) six months before mainshock, (c) six months after mainshock, and (d) one year after mainshock. The star denotes the mainshock epicenter.}
	\label{fig:spatial_dc}
\end{figure}

The spatial $D_c$ maps computed over the same two pre mainshock windows, one year (Figure~\ref{fig:spatial_dc}a) and six months (Figure~\ref{fig:spatial_dc}b), remain consistently close to 1.0 within the mainshock area and along the surrounding fault zones, suggesting that pre mainshock seismicity was concentrated along fault controlled linear structures rather than dispersed across broader planar or volumetric domains, indicating that deformation was primarily accommodated along the principal fault zones during the preparation phase. This is consistent with previous reports that fractal dimensions commonly decrease before large earthquakes as seismicity progressively localizes within the impending rupture zone \cite<e.g., >{wyss2004, fayou2010, ghosal2012, hamdache2019, yin2019, tiwari2022}, and with experimental evidence that fractal dimensions respond to the developing fracture process as failure approaches \citep{lei1999, chingtham2016, firoozfar2019, lei2019}.

In the one year window (Figure~\ref{fig:spatial_dc}a), a subtle reduction in $D_c$ develops near the epicenter, consistent with the progressive alignment and coalescence of small scale fractures along the future rupture plane \citep{oncel2002, papadakis2013}. Several localized patches with $D_c < 0.5$ occur elsewhere in the study area, indicating pronounced local clustering along individual fault segments and spatial heterogeneity in pre mainshock seismicity, while relatively higher $D_c$-values ($> 1.5$) occur mainly in the southeastern part of the study area, indicating a more dispersed pattern of seismicity in that sector. This organization becomes more pronounced in the six month window (Figure~\ref{fig:spatial_dc}b), where $D_c$ remains near 1.0 within the mainshock area while the southeastern sector exhibits a broader area of elevated values, implying that seismicity became more widely distributed in that part of the fault network as the mainshock approached.

\FloatBarrier

Following the mainshock (Figure~\ref{fig:spatial_dc}c, d), the $D_c$ field changes markedly. In both post mainshock windows, $D_c$-values in the mainshock region and its surroundings generally exceed 1.5, indicating that seismicity became distributed over a broader, more dispersed area than during the pre mainshock period, when values remained close to 1.0. This is consistent with aftershock activity expanding beyond the principal fault trace onto secondary faults and more distributed crustal deformation, activated by the redistribution of stress following the mainshock \cite<e.g., >{yadav2012, bayrak2013}. Six months after the mainshock, values above 1.5 become more widespread toward the southeastern part of the study area, and isolated low $D_c$ patches persist elsewhere, indicating that localized clustering continued on individual segments despite the overall spatial expansion of seismicity. One year after the mainshock, elevated $D_c$-values persist over an extensive fault network, with scattered low $D_c$ patches reflecting ongoing local clustering. The elevated $D_c$-values ($> 1.5$) in the southeastern region are consistent with the more volumetrically dispersed character expected of early aftershock distributions, in which aftershocks populate a broad volume surrounding the rupture plane as a spatially distributed network of secondary faults and fractures is activated \citep{Enescu2001, papadakis2013}. Overall, the post mainshock period is characterized by a spatially broader organization of seismicity than the pre mainshock stage, consistent with the redistribution of earthquake activity following the rupture. 

\subsection{Spatio-Temporal Depth Cross-Sections of $b$-Value and 3D $D_c$-Value Along Active Fault Zones}\label{sec:cross_sectional}
\begin{figure}[htbp]
	\centering
	\includegraphics[width=\textwidth]{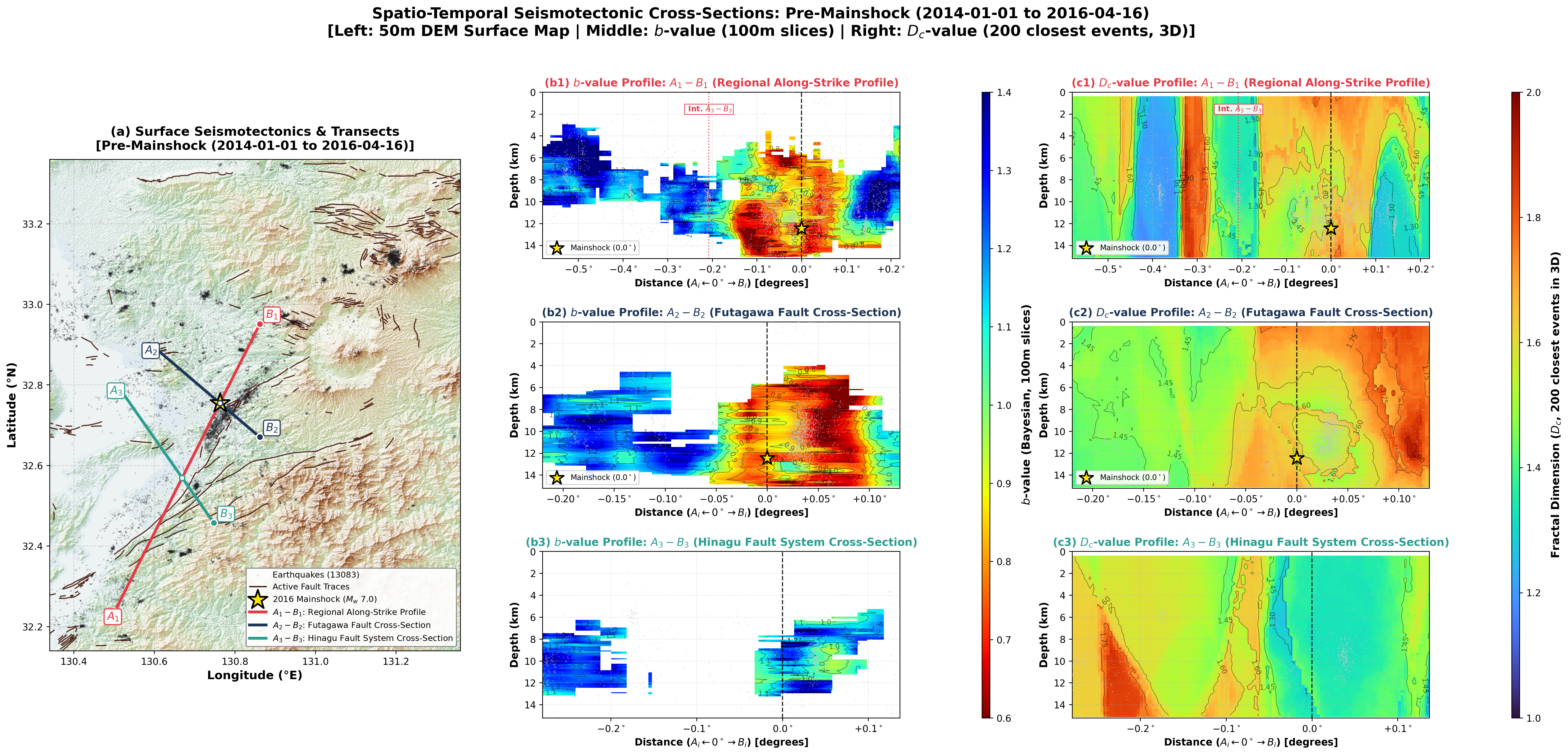}
	\caption{Pre-mainshock seismotectonic architecture and cross-sectional depth profiles (2014-01-01 to 2016-04-16). (a) Regional surface map on 50~m SRTM hillshade DEM showing active fault traces, pre-mainshock seismicity ($N=13,083$), 2016 mainshock ($M_w$~7.0, yellow star), and transects $A_1$--$B_1$ (red), $A_2$--$B_2$ (dark blue), and $A_3$--$B_3$ (teal). (b1--b3) Vertical depth cross-sections of Bayesian $b$-value ($100$~m vertical slices). (c1--c3) Corresponding 3D fractal dimension ($D_c$) computed from the 200 closest earthquakes in 3D Euclidean space. Horizontal axes denote angular distance in degrees relative to fault intersections ($0.0^\circ$).}
	\label{fig:tripanel_pre}
\end{figure}

\begin{figure}[htbp]
	\centering
	\includegraphics[width=\textwidth]{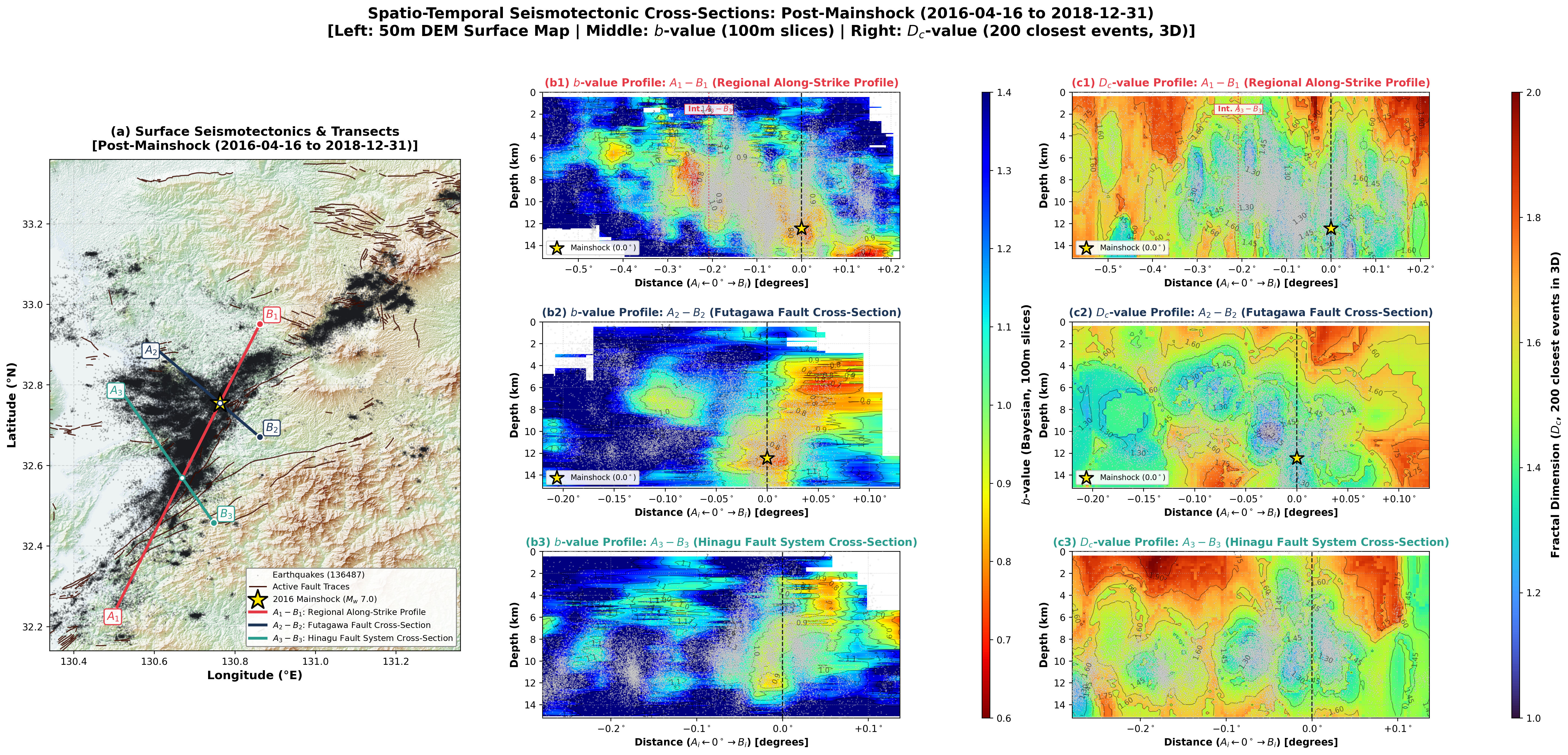}
	\caption{Post-mainshock seismotectonic architecture and cross-sectional depth profiles (2016-04-16 to 2018-12-31). (a) Regional surface map on 50~m SRTM hillshade DEM showing post-mainshock aftershock seismicity ($N=136,487$), active fault traces, and transects. (b1--b3) Vertical depth cross-sections of Bayesian $b$-value. (c1--c3) Corresponding 3D fractal dimension ($D_c$) computed from the 200 closest earthquakes in 3D Euclidean space.}
	\label{fig:tripanel_post}
\end{figure}

To resolve the depth-dependent stress state and 3D fracture architecture of the Kumamoto fault system, high-resolution cross-sectional profiles of both the Gutenberg--Richter $b$-value  and the fractal dimension $D_c$  were evaluated along three strategic transects ($A_1$--$B_1$, $A_2$--$B_2$, and $A_3$--$B_3$), shown in left panel of Figures~\ref{fig:tripanel_pre} and \ref{fig:tripanel_post}. To rigorously capture the coseismic stress perturbation and post-seismic structural reorganization, the analysis was performed separately for the pre-mainshock interval (2014-01-01 to 2016-04-16; Figure~\ref{fig:tripanel_pre}) and the post-mainshock interval (2016-04-16 to 2018-12-31; Figure~\ref{fig:tripanel_post}). 
All transects are plotted against angular distance in degrees relative to primary fault intersections ($0.0^\circ$ reference): $A_1$--$B_1$ follows the regional along-strike structural trend, intersecting $A_2$--$B_2$ strictly at the 2016 $M_w$~7.0 mainshock hypocenter ($130.7630^\circ\text{E}, 32.7545^\circ\text{N}$, depth 12.45~km; $0.0^\circ$) and intersecting the Hinagu fault transect ($A_3$--$B_3$) at $-0.21^\circ$; $A_2$--$B_2$ cuts orthogonally across the central Futagawa rupture plane through the mainshock nucleation point ($0.0^\circ$); and $A_3$--$B_3$ traverses the Hinagu fault system, centered at its intersection with $A_1$--$B_1$ ($0.0^\circ$).

\subsubsection{Pre-Mainshock Seismogenic Architecture: Asperity Locking and Precursory Damage}
During the pre-mainshock preparatory interval (Figure~\ref{fig:tripanel_pre}), the regional along-strike transect $A_1$--$B_1$ reveals an acute spatial contrast in differential stress and fracture complexity (Figures~\ref{fig:tripanel_pre}b1, c1). A pronounced, compact low $b$-value anomaly ($b \approx 0.60\text{--}0.75$) is tightly localized at seismogenic depths of 10 to 14~km directly beneath the mainshock epicenter ($-0.10^\circ$ to $+0.05^\circ$). Spatially coinciding with the mainshock nucleation point at 12.45~km depth, this zone represents a highly coupled, high-shear-stress asperity \citep{schorlemmer2005, scholz2015}. Conversely, the shallow-to-mid crust in the northwestern segment of $A_1$--$B_1$ (distances $-0.50^\circ$ to $-0.25^\circ$, depths 4 to 10~km) is characterized by elevated $b$-values ($b \approx 1.25\text{--}1.40$), indicating lower differential stress and thermal weakening proximal to the volcanic/geothermal domain. In the corresponding 3D $D_c$ profile (Figure~\ref{fig:tripanel_pre}c1), the high-stress nucleation core exhibits localized clustering ($D_c \approx 1.45\text{--}1.60$), whereas prominent high $D_c$ anomalies ($D_c \approx 1.75\text{--}1.95$) develop at the intersection with the Hinagu fault system ($-0.21^\circ$), diagnostic of a distributed, multi-scale fracture damage halo surrounding the locked asperity \citep{oncel2002}.
The fault-perpendicular profile $A_2$--$B_2$ (Figures~\ref{fig:tripanel_pre}b2, c2) clearly demonstrates vertical stress stratification across the Futagawa fault plane. At hypocentral depths of 10 to 14~km, a well-defined low $b$-value patch ($b < 0.75$) defines the locked core of the mainshock asperity, dipping slightly southeastward. The shallow overburden ($<6$~km depth) displays markedly higher $b$-values ($b > 1.20$), reflecting reduced lithostatic confining pressure and diffuse near-surface fracturing. The 3D $D_c$ values along $A_2$--$B_2$ exhibit moderate values ($D_c \approx 1.45\text{--}1.60$) along the deep primary fault plane, transitioning to elevated values ($D_c > 1.70$) in the surrounding block. Along transect $A_3$--$B_3$ (Figures~\ref{fig:tripanel_pre}b3, c3), moderate-to-low $b$-values ($b \approx 0.85\text{--}1.05$) are concentrated near the deep Hinagu--Futagawa junction ($0.0^\circ$ to $+0.08^\circ$), confirming elevated background stress concentration along the fault intersection prior to the main rupture.

\subsubsection{Post-Mainshock Reorganization: Coseismic Stress Release and Planar Rupture Localization}
Following the 2016 $M_w$~7.0 mainshock, the cross-sectional parameter distributions undergo extensive structural and stress reorganization across all three profiles (Figure~\ref{fig:tripanel_post}). Along the along-strike transect $A_1$--$B_1$ (Figures~\ref{fig:tripanel_post}b1, c1), the compact pre-seismic low $b$-value asperity undergoes significant stress relaxation, with $b$-values rising from pre-seismic values of $\sim 0.65$ to post-seismic values of $0.95\text{--}1.15$ around the mainshock hypocenter. Concurrently, localized low $b$-value corridors ($b \approx 0.75\text{--}0.85$) migrate toward the lateral rupture termination zones and deeper crustal roots ($\sim 14$~km depth), consistent with coseismic stress transfer and aftershock loading at the rupture margins \citep{tormann2015}. In the 3D $D_c$ profile (Figure~\ref{fig:tripanel_post}c1), sharp planar localization is manifested by distinct, low $D_c$ corridors ($D_c \approx 1.20\text{--}1.35$) tracing the primary slip interfaces, reflecting the concentration of aftershocks along discrete, planar rupture surfaces.
Across the Futagawa cross-section $A_2$--$B_2$ (Figures~\ref{fig:tripanel_post}b2, c2), the dense aftershock catalog delineates the inclined geometry of the primary Futagawa rupture. The core of the pre-mainshock asperity is replaced by moderate $b$-values ($b \approx 0.95\text{--}1.15$), with residual low $b$ patches preserved primarily along the deeper downdip boundary. The post-seismic 3D $D_c$ distribution clearly differentiates the highly localized, planar fault core ($D_c \approx 1.20\text{--}1.35$) from the broader, volumetric aftershock damage zone in the upper crust and hanging wall ($D_c > 1.65$). 
Along the Hinagu fault transect $A_3$--$B_3$ (Figures~\ref{fig:tripanel_post}b3, c3), dense aftershock activity reveals a continuous, steeply dipping seismogenic zone between depths of 4 and 14~km. A prominent low $b$-value corridor ($b \approx 0.75\text{--}0.95$) extends from $-0.05^\circ$ to $+0.10^\circ$, demonstrating sustained shear stress concentration and unclamping on unruptured and partially slipped segments of the Hinagu fault \citep{Kato2016, yoshida2017}. The corresponding 3D $D_c$ profile (Figure~\ref{fig:tripanel_post}c3) confirms strict planar localization ($D_c \approx 1.25\text{--}1.40$) along the Hinagu fault plane, bounded laterally by higher $D_c$ values indicative of secondary off-fault splay activation.
\FloatBarrier

\subsection{Integrated Multi Parameter Interpretation and Physical Implications}\label{sec:integrated}
Considered together, the evidence spanning the pre seismic preparation phase, the coseismic rupture, and the post seismic recovery of the 2016 Kumamoto $M_w$ 7.0 earthquake supports a coherent picture of how the seismogenic crust of central Kyushu evolved through the earthquake cycle. The $b$-value and $D_c$-value function as complementary gauges of crustal mechanical state, encoding distinct aspects of the stress field and fault geometry. In the pre seismic domain, the three parameters examined here, the $b$-value, the $D_c$-value, and the $Z$-value, trace a process of apparent fault locking and stress concentration that began roughly one to two years before the 16 April 2016 mainshock and became more pronounced in the final months. The $b$-value decline toward the future rupture zone is consistent with the progressive build up of shear stress on the seismogenic fault interface, following the inverse relationship between effective confining stress and the frequency magnitude scaling of small earthquakes \citep{schorlemmer2005, scholz2015}. The concurrent evolution of $D_c$, from a more distributed, moderate dimensional pattern to a spatially localized, lower dimensional configuration near the eventual epicenter, is consistent with the progressive coalescence of microfractures along the emerging rupture plane, reducing the spatial dimensionality of seismicity as the fault approaches failure \citep{oncel2002, papadakis2013}. These two signals are more diagnostic together than individually: as $b$ decreases on the locking fault, $D_c$ decreases near the epicenter while increasing on surrounding secondary structures, a spatially structured pattern consistent with stress transfer and Coulomb loading on off-fault structures \citep{king1994, stein1999, toda2011, toda2016science}. The seismic quiescence signal, while only reaching the adopted significance threshold in the longest, five month test window, adds a spatially coherent, independent line of evidence that is broadly consistent with the locking inference from $b$ and $D_c$ \citep{wyss1988, Enescu2001}. None of these three signals, considered alone, would justify a confident precursory interpretation; each carries substantial uncertainty and is subject to non tectonic contamination. Their joint, spatially overlapping behavior across the same fault zone and overlapping time windows is what makes the combined pattern noteworthy, though it should still be read as consistent with, rather than proof of, fault locking preceding the mainshock.

The post seismic evolution of the $b$-value and $D_c$-value fields shows a marked contrast with the pre seismic pattern. By six months after the mainshock, $b$-values in the vicinity of the rupture zone had increased substantially from their pre seismic minimum, in places exceeding 2.0, consistent with the coseismic stress drop reducing shear stress on and around the ruptured interface and shifting the earthquake size frequency distribution toward smaller events \citep{wiemer2002, Gulia2019}. Simultaneously, the $D_c$ field transitioned from the tightly organized, lower dimensional pre seismic configuration to a more broadly distributed aftershock pattern, consistent with activation of a wider network of secondary fractures and fault strands following the rupture \citep{Enescu2001, papadakis2013}. This post seismic state of elevated $b$ and elevated $D_c$ stands in contrast to the trajectory in the months before failure, during which both parameters had been declining toward their pre mainshock minimum. Over the six month and one year post seismic windows, both parameters evolve toward an intermediate state reflecting the superposition of ongoing aftershock decay, aseismic afterslip, and viscoelastic relaxation in the lower crust \citep{freed2005, perfettini2004}. The one year $b$-value map shows a differential recovery pattern, with elevated values persisting near the rupture zone while peripheral structures begin to reaccumulate tectonic stress, and the one year $D_c$ field shows a more heterogeneous pattern than either the pre seismic minimum or the immediate post seismic maximum. This gradual departure of the post seismic field from its pre seismic precursor is consistent with the expectation that fault strength recovery after large earthquakes can require years to decades through aseismic slip, pressure re equilibration, and mineral re cementation \citep{scholz1990, marone1998}, though the present catalog duration does not allow this recovery process to be tracked to completion.

The cross sectional profiles of $b$-value and $D_c$-value along the catalog integrated transects add a depth resolved structural dimension to this temporal picture. Across all three sections, the consistent identification of deep, low $b$ zones at depths of roughly 8 to 20 km, spatially coincident with the inferred fault patches of the Hinagu and Futagawa fault systems, indicates that the stress heterogeneity encoded in the $b$-value field has a structurally governed character over the catalog integration period. The depth dependent $D_c$ patterns, with higher values in the shallow, hydrothermally active crust and lower values in the deeper seismogenic zone, are consistent with a transition from a more complex, multiply faulted upper crust to a more geometrically simple, fault controlled seismicity pattern at depth. The spatial anticorrelation between low $b$ and high $D_c$ zones observed across the transects, in which the most strongly coupled fault domains are surrounded by more distributed, higher $D_c$ seismicity on secondary structures, mirrors the surface map pattern and is consistent with this being a genuine three dimensional structural feature of the seismogenic zone rather than an artifact of any single surface map. Because these cross sections aggregate the full catalog period, however, they are best read as evidence for a structurally persistent architecture rather than as direct, independent confirmation that this architecture is unchanged between the pre seismic and post seismic periods specifically.

Viewed as a whole, the dataset assembled here is consistent with a fault system that was progressively loaded toward failure over a period of one to two years, ruptured in April 2016, and has since been undergoing a measurable but, within the catalog duration examined here, incomplete process of mechanical recovery. The pre seismic convergence of low $b$, low $D_c$, and a spatially coherent, though only marginally significant, negative $Z$-value toward the rupture zone documents the loading phase with a degree of multi parameter agreement that adds confidence beyond any single parameter considered alone. The post seismic reversal and gradual relaxation of these same parameters documents the recovery phase and provides a baseline against which future parameter evolution in this region could be benchmarked. This pattern, in principle, is the kind of signature that could be tracked through routine catalog monitoring \citep{ozturk2018, wyss1988}. Its practical value for hazard assessment is nonetheless limited by the retrospective character of the present analysis: the anomalies reported here were identified with foreknowledge of the mainshock's location, timing, and magnitude, conditions that cannot be replicated in a prospective operational setting. Translating this kind of multi parameter framework into an operational monitoring tool would require prospective validation across a broader set of earthquake sequences, explicit uncertainty quantification at each analytical step, and a decision framework for converting continuous parameter monitoring into probabilistic hazard information \citep{Gulia2019, ozturk2018}. Within these limits, the evidence from the 2016 Kumamoto earthquake is consistent with the broader view that the seismicity record of well instrumented crustal fault systems can carry coherent, multi dimensional imprints of the mechanical state of the seismogenic crust across the seismic cycle, and that extracting these imprints with multi parameter methods of this kind is a scientifically grounded direction for further study.

\section{Conclusions}\label{sec:conclusions_new}
The 16 April 2016 $M_w$ 7.0 Kumamoto earthquake sequence provides an exceptionally well instrumented natural laboratory to investigate the geomechanical preparation, dynamic rupture, and post-seismic stress redistribution of an active intra-arc fault system across a complete earthquake cycle. By integrating high-resolution spatio-temporal analyses of the Gutenberg--Richter $b$-value, the three-dimensional hypocentral fractal dimension ($D_c$), and the standardized seismicity-rate anomaly ($Z$-value) from the dense Japan Meteorological Agency catalog, this study resolves a coherent, multi-parameter signature of crustal stress evolution and structural reorganization across the pre-seismic, coseismic, and post-seismic phases.

In the one to two years preceding catastrophic rupture, the seismogenic crust experienced progressive shear-stress concentration and hypocentral strain localization. The regional $b$-value underwent a pronounced, systematic decline from a background baseline of $b \approx 1.35 \pm 0.10$ to an anticipatory precursory minimum of $b \approx 0.59$, remaining depressed ($b \approx 0.73$--$0.79$) immediately prior to dynamic failure, consistent with intensifying shear stress on the locked fault interface. Concurrently, the three-dimensional hypocentral fractal dimension contracted from $D_c \approx 0.85 \pm 0.15$ to $D_c \approx 0.70$--$0.80$, indicating progressive coalescence of microfractures and localization of preparatory seismicity onto a quasi-linear, fault-parallel geometry enveloping the future nucleation zone. A spatially coherent negative $Z$-value anomaly ($Z \approx -1.6$ to $-2.0$), interpreted as a candidate seismic quiescence signal, developed along the Beppu--Shimabara graben; this anomaly strengthened with increasing integration time and approached, but only in the longest (five-month) test window reached, the adopted $|Z| > 2.0$ significance threshold. Individually, none of the three parameters would justify a confident precursory interpretation given their statistical uncertainty and susceptibility to non-tectonic contamination; it is their spatially and temporally overlapping behavior across the same fault volume that lends the composite pattern interpretive weight.

Pseudo-tomographic three-dimensional depth-sliced volumes and consecutive four-month spatio-temporal stacks further reveal that this precursory stress accumulation was vertically stratified and acutely concentrated in time. The shallow crust ($z < 5.0$~km) exhibited elevated background $b$-values ($1.20$--$1.40$), whereas a deep-seated low-$b$ core ($b \le 0.65$) was tightly focused at seismogenic depths of $10.0$ to $12.5$~km, directly enveloping the $M_w$ 7.0 nucleation hypocenter. Four-month non-overlapping temporal stacking demonstrated that this localized low-$b$ anomaly was absent throughout 2014 and 2015 and developed exclusively during the final four months preceding failure (February to April 2016), establishing that asperity locking was an acute, accelerating preparatory process rather than a static, long-lived tectonic feature.

Dynamic rupture and the associated coseismic shear-stress drop ($\Delta \tau \sim 1$--$10$~MPa) induced an immediate, dramatic reversal across all seismicity state variables. Within weeks of the mainshock, the $b$-value surged to $1.25$--$1.35$ and the fractal dimension expanded volumetrically to $D_c \approx 2.04$--$2.11$, while the $Z$-value time series reached its extreme value of $-1.50$, coincident with the onset of the aftershock-dominated activation phase. The high post-seismic fractal dimension ($D_c > 2.0$) confirms that aftershock deformation was not confined to a single rupture surface but was distributed volumetrically across the complex, conjugate fault network of the Futagawa and Hinagu segments. Over the subsequent 2.5 years, parameter trajectories exhibited a gradual recovery toward background levels ($b \approx 1.38 \pm 0.08$, $D_c \approx 1.60$--$1.85$, $Z$ rising to approximately $+1.14$), broadly tracking modified Omori-law aftershock decay and post-seismic viscoelastic relaxation \citep{omori1894, Utsu1995}.

High-resolution depth cross-sections ($A_1$--$B_1$, $A_2$--$B_2$, $A_3$--$B_3$) demonstrate that while the mainshock nucleation asperity underwent substantial stress relaxation ($b$ rising to $0.95$--$1.15$), post-seismic stress redistribution was highly heterogeneous. Two discrete, unrelaxed asperity patches, characterized by persistent low-to-moderate $b$-values ($b \approx 0.80$--$0.95$) and localized planar fracture geometry ($D_c \approx 1.20$--$1.35$), were identified at the unruptured southwestern termination of the Hinagu fault ($z = 10.0$--$17.5$~km) and near the northeastern rupture limit adjacent to the volcanic conduit of Mount Aso ($z = 7.5$--$12.5$~km). These zones coincide with areas of inferred positive Coulomb stress transfer ($\Delta\mathrm{CFS} > 0$) and structural fault-arrest barriers \citep{king1994, stein1999, toda2016science}, identifying crustal segments that retained elevated shear stress after the primary rupture had relaxed elsewhere. The consistent spatial anticorrelation between low-$b$, low-$D_c$ zones and their surrounding higher-$D_c$ domains across all three transects indicates that this structural pattern is a genuine three-dimensional feature of the fault system rather than an artifact of any single map projection.

We emphasize that the present study constitutes a retrospective diagnostic investigation of a single major historical event. Although the systematic, spatially and temporally coincident evolution of $b$, $D_c$, and $Z$ before, during, and after the mainshock provides empirical insight into the geomechanical physics of fault loading, asperity locking, and post-seismic recovery, retrospective analyses of this kind cannot by themselves establish operational forecasting capability: the anomalies reported here were identified with foreknowledge of the mainshock's location, timing, and magnitude, conditions that cannot be replicated in a prospective operational setting. Translating this multi-parameter framework into a genuine forecasting tool would require validation across a broader, independent set of earthquake sequences and tectonic environments, explicit quantification of false-alarm and failure-to-predict rates, and automated, real-time testing protocols free of retrospective tuning \cite<e.g., within the Collaboratory for the Study of Earthquake Predictability, CSEP; >{zechar2010, schorlemmer2018}, benchmarked against standard Poisson and Epidemic-Type Aftershock Sequence (ETAS) models using Molchan error diagrams, Receiver Operating Characteristic analysis, and Probability Gain scoring \citep{ogata1988, molchan1997, marzocchi2012}. Within these limits, the 2016 Kumamoto sequence is consistent with the broader view that densely instrumented crustal fault systems can retain coherent, multi-dimensional imprints of their mechanical state across the seismic cycle, and that continued refinement of joint seismicity-parameter frameworks represents a scientifically grounded direction for future study.

\section*{Open Research Section}
The earthquake catalog data used in this study were obtained from the Japan Meteorological Agency (JMA) unified catalog at \url{https://www.jma.go.jp/jma/index.html}. Calculations were performed using Python packages and the ZMAP software package \cite{Wiemer2001}.

\section*{Conflict of Interest declaration}
The authors declare there are no conflicts of interest for this manuscript.

\acknowledgments
The authors thank the Japan Meteorological Agency for providing the high-quality earthquake catalog. M.H.M. was supported by the Atmosphere and Ocean Research Institute (AORI), the University of Tokyo.

\bibliography{mybibfile}

\end{document}